\documentclass[11pt,reqno]{amsart}

\usepackage{amssymb}
\usepackage{amsthm}
\usepackage{amsmath}
\usepackage{stmaryrd}
\usepackage{mathtools}
\usepackage{url}
\usepackage{hyperref}
\usepackage[dvips]{color}
\usepackage{tikz-cd}
\usepackage[all]{xy}
\SelectTips{cm}{}

\allowdisplaybreaks[4]

\usepackage{mathrsfs}
\let\mathcal\mathscr
\usepackage[mathscr]{eucal}
\usepackage{tabularray} 

\newcommand*{\pd}[2]{\mathchoice{\frac{\partial#1}{\partial#2}}
  {\partial#1/\partial#2}{\partial#1/\partial#2}
  {\partial#1/\partial#2}}

\let\phi=\varphi
\let\kappa=\varkappa
\let\epsilon=\varepsilon

\DeclareMathOperator{\sym}{sym}

\newcommand*{\jac}[2]{\left\{ #1,#2  \right\}}

\newcommand\restr[2]{{
  \left.\kern-\nulldelimiterspace 
  #1 
  \littletaller 
  \right|_{#2} 
  }}

\newcommand{\littletaller}{\mathchoice{\vphantom{\big|}}{}{}{}}

\newdir{ >}{{}*!/-5pt/\dir{>}}

\theoremstyle{theorem}

\newtheorem{proposition}{Proposition}
\newtheorem{corollary}{Corollary}

\newtheorem{conj}{Conjecture}

\theoremstyle{definition}

\theoremstyle{remark}
\newtheorem{remark}{Remark}

\usepackage{mathrsfs}
\let\mathcal\mathscr
\usepackage[mathscr]{eucal}
\newcommand{\cprime}{\/{\mathsurround=0pt$'$}}

\author{Ji\v rina Jahnov\'a, Petr~Voj{\v{c}}{\'{a}}k}
\address{Mathematical Institute, Silesian University in Opava, Na Rybn\'{\i}\v{c}ku 1, 746 01 Opava, Czech Republic}
\email{Jirina.Jahnova@math.slu.cz, Petr.Vojcak@math.slu.cz} 
\title[$\Bbb Z$-graded Lie algebras of full-fledged nonlocal symmetries]{A straightforward construction of $\Bbb Z$-graded Lie algebras of full-fledged nonlocal symmetries via recursion operators}

\begin{document}

\begin{abstract}
We consider the reduced quasi-classical self-dual Yang-Mills equation (rYME) and two recently found (Jahnov\'{a} and Voj\v{c}\'{a}k, 2024) invertible recursion operators $\mathcal{R}^q$ and $\mathcal{R}^m$ for its full-fledged (in a given differential covering) nonlocal symmetries.  We introduce a $\mathbb{Z}$-grading on the Lie algebra $\mathrm{sym}_{\mathcal{L}}^{\tau^W} (\mathcal E)$ of all nonlocal Laurent polynomial symmetries of the rYME and prove that both the operators $\mathcal{R}^q$ and $\mathcal{R}^m$ are $\mathbb{Z}$-graded automorphisms of the underlying vector space on the set $\mathrm{sym}_{\mathcal{L}}^{\tau^W} (\mathcal E)$. This \textit{inter alia} implies that all its vector subspaces formed by all homogeneous elements of a given fixed  degree (i.e. a weight in the context below) are mutually isomorphic, and thus each of them can be uniquely reconstructed from the vector space of all homogeneous symmetries of the zero degree. To the best of our knowledge, such a result is unparalleled in the current body of literature. The obtained results are used for the construction of a Lie subalgebra $V$ of $\mathrm{sym}_{\mathcal{L}}^{\tau^W} (\mathcal E)$ which contains all known to us nonlocal Laurent polynomial symmetries of the rYME. The Lie algebra $V$ is subsequently described as the linear span of the orbits of a set of selected zero-weight symmetries - we refer to them as to the seed generators of $V$. Further, we study the hierarchies of symmetries related to the seed generators under the action of the group of recursion operators generated by $\mathcal{R}^q$ and $\mathcal{R}^m$. Finally, the linear dependence/independence of the (sub)set of generators of $V$ is discussed.
\end{abstract}

\subjclass[2020]{35B06, 17B70}
\keywords{Weights, $\Bbb Z$-graded Lie algebras, nonlocal symmetries, recursion operators}
\maketitle

\section*{Introduction}
\label{sec:introduction}
It is well known that infinite-dimensional Lie algebras of nonlocal symmetries play an important role in the theory of integrable systems and provide us with a useful tool for the study of the latter (see e.g. \cite{kra1}, \cite{kra3}). Therefore, it is no wonder that nonlocal symmetries are currently a subject of great interest in many studies, see e.g. \cite{Bar, duar, jin, KMV, KrasVoj, Mor-Ser, shag, vin}. However, the task to find and describe this Lie algebra is by no means easy.
Nonlocal symmetries are typically sought by solving the equations that describe their first components, often referred to as shadows. Once a shadow is identified, one can try to search for further components, though not every shadow corresponds to a full nonlocal symmetry, see e.g. \cite{boch}. Just like in the case of local symmetries, recursion operators could be an important tool in the search for nonlocal symmetries as well. However, the traditionally used recursion operators usually have the form of B\"{a}cklund auto-transformations, which are a kind of differential relation between two shadows of the same equation, see e.g. \cite{boch, kramor, KMV, kra1, kra2, Marvan95, Mor1, Serg, voj} and references therein for more details.  But the drawback of such recursion operators is obvious: if we aim to construct the Lie algebra of full-fledged nonlocal symmetries (recall that the set of shadows does not form any Lie algebra), we also repeatedly encounter the aforementioned necessity of lifting shadows of symmetries to their full-fledged forms.
This disadvantage can be overcome by using recursion operators for full-fledged nonlocal symmetries, i.e. mappings that directly map full-fledged nonlocal symmetries to other ones. Such an approach has been recently presented in \cite{jahn_voj} where 
mutually commuting recursion operators $\mathcal{R}^q$ and $\mathcal{R}^m$ (and in fact a whole group of recursion operators generated by $\mathcal{R}^q$ and $\mathcal{R}^m$) for full-fledged nonlocal symmetries of the reduced-classical self-dual Yang-Mills equation (abbreviated as the rYME)
\begin{equation}
\label{yme}
\mathcal{E}: u_{yz} = u_{tx} - u_z u_{xx}+u_xu_{xz}
\end{equation}
were found. We recall that the rYME \eqref{yme} was first introduced in \cite{fer}, subsequenty studied e.g. in \cite{kramor}, \cite{KrugMor2016}, \cite{mor}, \cite{zhang}, and it also arises in the context of the classification  \cite{DFKN} of the integrable four-dimensional systems associated with sixfolds in \textbf{Gr}(4,6).

The present paper focuses on further study of the recursion operators $\mathcal{R}^q$ and $\mathcal{R}^m$. We work in the differential covering $\tau^W$ of the rYME \eqref{yme} that was constructed as the Whitney product of three differential coverings obtained from the expansion of the two-parameter Lax pair, see \cite{jahn_voj} for more detail.
We narrow our interest solely to those $\tau^W$-symmetries, whose generating functions are sequences of differential functions lying in the class of Laurent polynomials of internal coordinates on the covering equation, and introduce a $\mathbb{Z}$-grading on the Lie algebra $\mathrm{sym}_{\mathcal{L}}^{{\tau}^W}(\mathcal{E})$ of these $\tau^W$-symmetries. This allows us to study further properties of the recursion operators in question, in particular, we show that all of them are graded homogeneous automorphisms of the vector space  $\mathrm{sym}_{\mathcal{L}}^{{\tau}^W}(\mathcal{E})$. The obtained results \textit{inter alia} imply that all the vector subspaces of the vector space $\mathrm{sym}_{\mathcal{L}}^{{\tau}^W}(\mathcal{E})$  formed by all homogeneous elements of a given fixed degree (weight) are mutually isomorphic, and thus each of them can be uniquely reconstructed from the vector space of  all homogeneous symmetries of the zero degree. Our goal is to describe the infinite-dimensional Lie subalgebra $V \subset \mathrm{sym}_{\mathcal{L}}^{{\tau}^W}(\mathcal{E})$ of all known to us symmetries as the linear span of the elements of the orbits of several selected $\tau^W$-symmetries (the so-called seed generators of V) under the iterative application of the operators $\mathcal{R}^q,\ \mathcal{R}^m$ and their inverses, and thus to describe the generators of $V$ by a recursive formula.

The structure of the paper is as follows. In Section 1, we recall a few necessary notions and facts from the geometrical theory of PDEs, see e.g. \cite{boch}, \cite{kra1}, \cite{kra2} for more detailed exposition of the latter. Section 2 is devoted to the problem of introducing $\mathbb{Z}$-grading on the Lie algebra of polynomial symmetries of a general PDE.  In Section 3, we briefly summarize the main results concerning the rYME \eqref{yme}, which were obtained in \cite{jahn_voj}. In Section 4, we use the general construction described in Section 2 and introduce a $\mathbb{Z}$-grading on the Lie algebra $\mathrm{sym}_{\mathcal{L}}^{\tau^W}(\mathcal{E})$ of full-fledged nonlocal Laurent polynomial symmetries of the rYME \eqref{yme}. We prove that both the basic recursion operators $\mathcal{R}^q$ and $\mathcal{R}^m$ are graded homogeneous $\mathbb{R}$-linear transformations of $\mathrm{sym}_{\mathcal{L}}^{\tau^W}(\mathcal{E})$, the first one of degree $-1$, the second one of degree $0$. From this, among other things, it can be inferred that the orbit $\langle\mathcal{R}^q\rangle\cdot P$ of a non-zero homogeneous symmetry $P$ is a linearly independent set that contains infinitely many symmetries.
In Section 5, we describe the Lie algebra $V$ of all known to us $\tau^W_\mathcal L$-symmetries of the rYME \eqref{yme} as the linear span of the elements of the $\langle\mathcal{R}^q,\mathcal{R}^m\rangle$-orbits of a set of selected $\tau^W_\mathcal L$-symmetries - the seed generators of $V$. Hence, each symmetry can be computed via application of the full-fledged recursion operators and summation, without any problems concerning lifting of shadows. 
We further discuss the linear dependence/independence of the generator set of  $V.$
Finally, we consider the action of the group $\langle\mathcal{R}^m,\mathcal{R}^q\rangle$ on the set $\mathrm{sym}_{\mathcal{L}}^{\tau^W}(\mathcal{E})$ and give a partial description of the orbits of the seed generators.

\section{Preliminaries}
Below we work in the jet space $J^{\infty}(\pi)$ where $\pi:\mathbb{R}^m\times\mathbb{R}^n\to\mathbb{R}^n$ is the trivial bundle with coordinates $x^i,u^j,u_{\sigma}^j$, $i=1,\dots,n$, $j=1,\dots,m$:  here $x^i$ stand for independent variables, $u^j$ for the dependent variables, and $u^j_\sigma$ stands for the derivative of the dependent variable $u^j$ with respect to the independent variables stated in the multiindex $\sigma$ ($\sigma$ are symmetric multi-indices of arbitrary finite length that consist of the integers $1,\dots,n$). The space $J^{\infty}(\pi)$ is equipped with \textit{the total derivative operators}
$$D_{x^i}\equiv D_i = \frac{\partial}{\partial x^i} + \sum \limits_{j,\sigma} u_{\sigma i}^j \frac{\partial}{\partial u_\sigma^j},\quad i=1,\dots,n,$$
that generate the so-called \textit{Cartan distribution} $\mathcal{C}$ on it. In what follows, $D_{\sigma}$ denotes the composition of the total derivative operators with respect to the independent variables stated in the multiindex $\sigma$.

Let $\mathcal{F}(\pi)$ denote the $\mathbb{R}$-algebra of all smooth functions on $J^{\infty}(\pi)$ that depend explicitly on only finite number of the variables. These functions are called the \textit{differential functions}. Let 
\begin{eqnarray}
\label{rovnice}
F=(F^1,F^2,\dots, F^r)=0, \quad F^i\in\mathcal{F}(\pi),\; i=1,\ldots,r,
\end{eqnarray} 
be the system of differential equations given by the $r$-tuple $F$ of differential functions. The submanifold 
\begin{equation}
\label{eqE}
\mathcal{E}\equiv \mathcal{E}_F=\left\{\theta\in J^{\infty}(\pi)\ |\ D_{\sigma}(F^{\alpha})(\theta)=0\ \forall \alpha,\sigma\right\}\subset J^{\infty}(\pi)
\end{equation}
 equipped with the Cartan distribution restricted to $\mathcal{E}$ ( $\restr{\mathcal{C}}{{\mathcal{E}_F}}$ is spanned by the vector fields $\bar {D_i}=\restr{D_{i}}{{\mathcal{E}_F}}$
restricted to $\mathcal{E}$) is then called \textit{the infinitely prolonged equation}.

The \textit{smooth functions on $\mathcal{E}$} are defined to be the elements of the factor algebra $\mathcal{F}({\mathcal{E}})=\mathcal{F}(\pi)/I({\mathcal{E}})$, where 
$$I({\mathcal{E}})=\left\{\psi\in\mathcal{F}(\pi)\ |\ \psi=\sum_{\rho,\sigma}\psi_{\rho,\sigma}{D}_{\sigma}(F_{\rho}),\psi_{\rho,\sigma}\in\mathcal{F}(\pi)\right\}\subset\mathcal{F}(\pi)$$ 
is the ideal of ${\mathcal{E}}$ (see \cite{boch} for more details). Thus, a smooth function on $\mathcal{E}$  (working in the coordinates of $J^{\infty}({\pi})$) takes the form of a coset ${\psi}+I({\mathcal{E}})$, where $\psi\in\mathcal{F}(\pi)$.

Suppose that it is possible to resolve the equations $F^i=0$, $i=1,\ldots,r$, with respect to certain partial derivatives, and thus to choose and fix a set $\mathbb{I}$ of \textit{internal coordinates} on $\mathcal{E}$. 
Then each coset $\bar\psi=\psi+I({\mathcal{E}})\in\mathcal{F}(\mathcal{E})$ shrinks to only one function $\bar\psi_{\mathrm{int}}$, which is the function $\psi$ expressed in the internal coordinates on ${\mathcal{E}}$.

\begin{remark}
Note that the resulting function $\bar{\psi}_{\mathrm{int}}$ may be (considered as the function of internal variables) smooth  only locally. Therefore, working in the internal coordinates, the algebra $\mathcal{F}(\mathcal{E})$ can be considered to be the algebra of  functions that are locally smooth as the functions of the internal coordinates on $\mathcal{E}$.
\end{remark}

A \textit{(higher infinitesimal) symmetry} of the equation ${\mathcal E_F}$ is  an evolutionary vector field
\begin{equation}
\label{defsym}
\mathbf E_{\Phi} =\sum
\limits_{j,\sigma}
\bar D_{\sigma}(\phi_j)\frac{\partial}{\partial u_\sigma^j}, \quad \varphi_j\in\mathcal{F}(\mathcal{E}),
\end{equation}
where $\Phi=(\phi_1,\phi_2, \ldots,\phi_m)$ satisfies the condition
\begin{equation}
\label{lin_sym_cond}
\bar{\ell}_F(\Phi)\equiv\restr{ \ell_F}{\mathcal{E}}(\Phi)= 0.
\end{equation}
Here
$$\ell_F=\begin{pmatrix}
\sum_\sigma \frac{\partial F^1}{\partial u_\sigma^1}D_\sigma & \sum_\sigma \frac{\partial F^1}{\partial u_\sigma^2}D_\sigma & \ldots \\[3mm]
\sum_\sigma \frac{\partial F^2}{\partial u_\sigma^1}D_\sigma & \sum_\sigma \frac{\partial F^2}{\partial u_\sigma^2}D_\sigma & \ldots\\
\vdots & \vdots & \ddots
\end{pmatrix}
$$
denotes the \textit{operator of linearization of $F$}. The vector-valued differential function $\Phi\in \mathcal{F}\,^m(\mathcal{E})$ is called the \textit{generating function of the symmetry} $\mathbf E_{\Phi}$.  

The set of all symmetries of the equation $\mathcal E$ equipped with the commutator forms a Lie algebra $\sym(\mathcal E)$. It is isomorphic to the Lie algebra of all  $m$-tuples $\Phi\in \mathcal{F}\,^m(\mathcal{E})$ that are the solutions to \eqref{lin_sym_cond}, the Lie algebra structure being given by the Jacobi bracket $\jac{\cdot}{\cdot}$ defined by the formula
\begin{equation}
\label{sec1:2}
\jac{\Phi_1}{\Phi_2} = \mathbf E_{\Phi_1}(\Phi_2)-\mathbf E_{\Phi_2}(\Phi_1).
\end{equation}
Due to this isomorphism we do not distinguish between the symmetries and their generating functions.

A \textit{recursion operator} for symmetries of $\mathcal{E}$ (in the classical sense of Olver, see e.g. \cite{Olver93}) is an $\mathbb{R}$-linear mapping $\mathcal{R}:\mathcal{F}\,^m(\mathcal{E})\to\mathcal{F}\,^m(\mathcal{E})$ with the property that whenever $\Phi$ is a generating function of a symmetry of $\mathcal{E}$, so is $\mathcal{R}(\Phi)$.

Let $w^1,\dots,w^l$ (where $l$ possibly may be infinite) be new unknown functions of the variable $x=(x^1,\dots,x^n)$, let $w^k_{\sigma}$ stand for the derivatives of the function $w^k$ with respect to the variables stated in the multi-index $\sigma$. The \textit{differential covering} of the equation $\mathcal{E}$ is the projection $\tau:\tilde{\mathcal{E}}\to\mathcal{E}, (x, u_{\sigma}^j,w_{\sigma}^{\eta})\mapsto(x, u_{\sigma}^j)$, where the covering space $\tilde{\mathcal{E}}$ is the infinitely prolonged equation $\tilde{\mathcal{E}}\subset J^{\infty}(\tilde\pi)$ given by the system of differential equations
\begin{equation}
\label{coveq}
\begin{aligned}
F^s(x,u_{\sigma}^j)&=0,\\
\frac{\partial w^{\eta}}{\partial x^i}&=X_i^\eta(x,u_{\sigma}^j,w^{\gamma}), \quad i=1,\ldots,n,\quad j=1,\dots,m,\quad \eta, \gamma=1,\ldots,l,\quad s=1,\ldots,r,
\end{aligned}
\end{equation}
where the compatibility conditions for \ref{coveq} are the differential consequences of the equation \ref{rovnice}.

The symmetries of the equation $\tilde{\mathcal E}$ are said to be \textit{nonlocal $\tau$-symmetries of the equation $\mathcal E$}. It holds (see e.g. \cite{boch}, \cite{kra1}, \cite{kra2}) that they take the form
\begin{equation}
\label{evol-non-sym}
\mathbf{E}_{(\Phi,\Psi)}=\sum \limits_{\sigma,j} \tilde D_{\sigma}(\phi_j)\frac{\partial}{\partial u_\sigma^j} +\sum \limits_\eta \psi^\eta \pd{}{w^\eta},
\end{equation}
where $\Phi=(\phi_1, \phi_2,\ldots,\varphi_m)$ and $\Psi=(\psi_1, \psi_2, \ldots,\psi_l)$ are vector-valued functions on $\tilde{\mathcal E}$ satisfying the system
\begin{align}
\label{sec1:4}
\tilde \ell_{\mathcal E}(\Phi)&=0,\\[2mm]
\label{sec1:5}
\tilde D_i(\psi^\eta) & = \tilde \ell_{X_i^\eta}(\Phi) + \sum \limits_\nu \pd{X_i^\eta}{w^\nu}\psi^\nu,
\end{align}
and $\tilde{D}_{\sigma}$, resp. ${\tilde{\ell}}_{\mathcal{E}}$, denote the natural lift of the differential operator  $D_{\sigma}$, resp. ${\ell}_{\mathcal{E}}$, from $\mathcal E$ to $\tilde{\mathcal E}$. Solutions of the equation \eqref{sec1:4} are called \textit{shadows of nonlocal $\tau$-symmetries}. Nonlocal $\tau$-symmetries with the trivial shadow are called \textit{invisible $\tau$-symmetries}.
\begin{remark}
Even though the set of all shadows of nonlocal symmetries does not carry the Lie algebra structure, there are some authors who call them to be nonlocal symmetries. Therefore, in order to avoid any confusion, we sometimes call the nonlocal $\tau$-symmetries as the \textit{full-fledged nonlocal $\tau$-symmetries.}
\end{remark}

\section{The $\mathbb{Z}$-grading on the Lie algebra of (nonlocal) symmetries}
\label{grading}
If it is possible to suitably introduce them (typically if the equation ${\mathcal E}$  admits at least one scaling symmetry) weights ($\Bbb Z$-grading) provide us with a very useful tool in analysis of Lie algebras structures related to the given equation $\mathcal E$, see. e.g. \cite{BKMV, kra2} and references therein for more details. Since the concept of the weights plays a very important role in our forthcoming exposition, we devote this section to a systematic description of how such grading exactly arise on the Lie algebra of (nonlocal) symmetries (or of the generating functions of the latter). As the basis of the $\mathbb{Z}$-grading construction is formed by the so-called weight setting, we start with its definition. 

 An arbitrary map
  \begin{equation}
 \label{weight}
 |\cdot|:\left\{x^i,u^j\ | \ i=1,\ldots,k,\ j\geq 1 \right\}\to\mathbb{Z}
 \end{equation} 
on the set of symbols denoting the independent and dependent variables, will be referred to as a \textit{weight setting}. 

Having fixed an arbitrary weight setting, we are subsequently able to introduce a $\mathbb{Z}$-grading on several algebraic structures related to the jet space $J^\infty(\pi)$.

\subsection{The $\mathbb{Z}$-grading on the ring $\mathcal{L}(\pi)$ of the Laurent polynomial functions on $J^{\infty}(\pi)$}
The assignment of the weight to an arbitrary nonzero monomial in the jet coordinates $x^i,u_{\sigma}^j$  is carried out according to the following rules (the symbols $a$ and $b$ stand for arbitrary jet coordinates here):
\begin{itemize}
 \item[a)] For an arbitrary multiindex $\sigma=(\sigma_1,\dots,\sigma_k)$ we define the weight of $u^j_{\sigma}$ as
\begin{equation}
\label{wmon_1}
\left| u^j_{\sigma} \right| := |u^j|-|x^{\sigma_1}|-\ldots - |x^{\sigma_k}|,
\end{equation}
thus, for example, $|u^j_{x^1x^1x^3}|\equiv |u^j_{(1,1,3)}|=|u^j|-|x^1|-|x^1|-|x^3|$.
 \item[b)] The weight of a formal $\gamma$-power of the symbol $a$ is defined as
\begin{equation}
\label{wmon_2}
|a^{\gamma}|:=\gamma\cdot |a|\ \mbox{for\ all}\ \gamma\in\mathbb{Z}.
\end{equation}
  \item[c)] The weight of a (formal) product of two symbols $a^{\gamma_1}$ and $b^{\gamma_2}$, $\gamma_1,\gamma_2\in\mathbb{Z}$, is defined as
\begin{equation}
\label{wmon_3}
|a^{\gamma_1}b^{\gamma_2}|:=|a^{\gamma_1}|+|b^{\gamma_2}|.
\end{equation}
\end{itemize}

Subsequently, a nonzero Laurent polynomial\footnote{A \textit{Laurent polynomial in the indeterminates} $X_1,\dots X_s$ with real coefficients is a polynomial in the indeterminates $X_1,X_1^{-1},\dots,X_s,X_s^{-1}$ with real coefficients. Whenever we talk about functions that are Laurent polynomials of the coordinates on $J^{\infty}(\pi)$, we always implicitly assume that they depend on only finitely many variables.} $p\in\mathcal{L}(\pi)$ is said to be \textit{homogeneous of the weight $|p|=k$} if all of its nonzero terms have the same weight equal to $k$. Note that the weight of a zero polynomial is not well-defined: \textit{the zero polynomial} is considered to be homogeneous of an arbitrary weight.

The set $\mathcal{L}_k(\pi)$ of all homogeneous Laurent polynomials of the weight $k$ carries the real vector space structure and it holds $\mathcal{L}(\pi)=\bigoplus_{k\in\mathbb{Z}}\mathcal{L}_k(\pi)$. Moreover, for homogeneous polynomials $p_1$ and $p_2$ we have $|p_1p_2|=|p_1|+|p_2|$, thus $\mathcal{L}(\pi)$ is a $\Bbb Z$-graded ring.

\subsection{The $\mathbb{Z}$-graded $\mathcal{L}(\pi)$-module structure and  the $\mathbb{Z}$-graded  Lie algebra structure on the set $\mathrm{Vect}_{\mathcal{L}(\pi)}(\pi)$ of vector fields on $J^{\infty}(\pi)$ with Laurent polynomial coefficients}
Let $a$ denote an arbitrary coordinate on $J^{\infty}(\pi)$ and let $f\in\mathcal{L}_k(\pi)$ be a homogeneous Laurent polynomial. The weight of the vector field $\mathbf{v}=f\frac{\partial}{\partial a}$ is defined as the degree of the graded homomorphism $\mathcal{L}(\pi)\to\mathcal{L}(\pi)$, $p\mapsto f\partial p/\partial a$, provided by $\mathbf{v}$,
that is
$$|\mathbf{v}|:=|f|-|a|.$$

A general vector field on $J^{\infty}(\pi)$ is said to be \textit{homogeneous of the weight $k$}, if all the nonzero summands have the weight $k$.  In particular, one can simply see that the total derivatives operators $D_{x^i}$ are homogeneous of the weights $|D_{x^i}| = -|x^i|$. The trivial vector field is considered to be homogeneous of arbitrary weight.

Note that given a homogeneous function $g\in\mathcal{L}(\pi)$ we have $|g\mathbf{v}|=|g|+|\mathbf{v}|$, thus $\mathrm{Vect}_{\mathcal{L}(\pi)}(\pi)$ becomes a $\mathbb{Z}$-graded $\mathcal{L}(\pi)$-module.

Moreover, it can be verified that the commutator $[\mathbf{v}_1,\mathbf{v}_2]$ of two homogeneous vector fields $\mathbf{v}_1$ and $\mathbf{v}_2$ is a homogeneous vector field with the weight $|[\mathbf{v}_1,\mathbf{v}_2]|=|\mathbf{v}_1|+|\mathbf{v}_2|$, thus $\mathrm{Vect}_{\mathcal{L}(\pi)}(\pi)=\bigoplus_{k\in\mathbb{Z}}\mathrm{Vect}_{{\mathcal{L}(\pi)},k}(\pi)$ is a $\Bbb Z$-graded Lie algebra.
 
\subsection{The $\mathbb{Z}$-graded $\mathcal{L}(\pi)$-module structure and  the $\mathbb{Z}$-graded  Lie algebra structure on the set $\mathcal{L}^m(\pi)$ of $m$-tuples of Laurent polynomial differential functions on $J^{\infty}(\pi)$ }
Firstly, note that the Lie subalgebra $\mathrm{Vect}_{\mathcal{L}(\pi),\mathrm{ev}}(\pi)\subset\mathrm{Vect}_{\mathcal{L}(\pi)}(\pi)$ of all evolutionary vector fields on $J^{\infty}(\pi)$ inherits the $\mathbb{Z}$-grading from the Lie algebra  $\mathrm{Vect}_{\mathcal{L}(\pi)}(\pi)$. 
Thus, the $\mathbb{Z}$-graded Lie algebra structure on  $\mathcal{L}^m(\pi)$  is naturally induced by means of the isomorphism between the Lie algebra $\mathrm{Vect}_{\mathcal{L}(\pi),\mathrm{ev}}(\pi)$ and the Lie algebra  $\mathcal{L}^m(\pi)$ (equipped with the Jacobi bracket).

Therefore, 
a function $\Phi=(\phi_1,\dots,\phi_m) \in \mathcal{L}^m(\pi)$ is \textit{homogeneous of the weight $|\Phi|=k$} if and only if the evolutionary vector field
$\mathbf{v}=\mathbf{E}_{\Phi}=\sum_{j,\sigma}D_{\sigma}(\phi_j)\frac{\partial}{\partial u_{\sigma}^j}$ is homogeneous of the weight $k$, that is, if and only if all the nonzero components of $\Phi$ are homogeneous and satisfy
\begin{equation}
\label{weight_evol_field}
|\phi_j|=k+|u^j|\ \ \mbox{for all} \ j=1,\dots.,m.
\end{equation}
We denote the vector space of all homogeneous $m$-tuples of Laurent polynomial functions on $J^{\infty}(\pi)$ of the weight $k$ as $\mathcal{L}^m_k(\pi)$.  Hence, we obviously have $\mathcal{L}^m(\pi)=\bigoplus_{k\in\mathbb{Z}}\mathcal{L}^m_k(\pi)$. 

Further, given a homogeneous $m$-tuple  $\Phi=(\phi_1,\dots,\phi_m) \in \mathcal{L}^m(\pi)$ and a homogeneous function $\psi\in\mathcal{L}(\pi)$, we can see that
$\psi\cdot\Phi=(\psi\phi_1,\dots,\psi\phi_m)$ is again a  homogeneous $m$-tuple, its weight being $|\psi\Phi|=|\psi|+|\Phi|$. Hence, $\mathcal{L}^m(\pi)=\bigoplus_{k\in\mathbb{Z}}\mathcal{L}^m_k(\pi)$ carries also the $\mathbb{Z}$-graded $\mathcal{L}(\pi)$-module structure.

\begin{remark}
In what follows, it is absolutely essential to distinguish between the homogeneous $m$-tuple of polynomials and the $m$-tuple of homogeneous polynomials. While the first notion implies that all polynomials in question are homogeneous and, moreover, their weights satisfy the condition \eqref{weight_evol_field}, the second one does not require these conditions to be satisfied.
\end{remark}

\begin{remark}
Note that the rules \eqref{wmon_1} - \eqref{wmon_3} provide us also with the $\mathbb{Z}$-grading on the ring $\mathcal{P}(\pi)$ of all polynomial functions on $J^{\infty}(\pi)$, which is the subring of the ring $\mathcal{L}(\pi)$. Hence, in analogy with the constructions made above, we are given with the $\Bbb Z$-graded $\mathcal{P}(\pi)$-module and  the $\mathbb{Z}$-graded Lie algebra structures on the set  $\mathrm{Vect}_{\mathcal{P}(\pi)}(\pi)$ of all vector fields on $J^{\infty}(\pi)$ with polynomial coefficients,  as well as on the set $\mathcal{P}^m(\pi)$ of all $m$-tuples of polynomial functions on $J^{\infty}(\pi)$.
\end{remark}
 Next, let us restrict our consideration only to the $\Bbb Z$-graded ring $\mathcal{P}(\pi)$ of polynomial differential functions on $J^{\infty}(\pi)$ and the related $\mathbb{Z}$-graded structures.
 
\subsection{The $\mathbb{Z}$-graded $\mathcal{P}(\mathcal{E})$-module structure and  the $\mathbb{Z}$-graded  Lie algebra structure on the set ${\mathcal{P}^m(\mathcal{E}})$ of all $m$-tuples of polynomial functions on an equation $\mathcal{E}$}
Consider an equation $\mathcal{E}=\mathcal{E}_F\subset J^{\infty}(\pi)$ defined by an $r$-tuple $F=(F^{1},\dots, F^r)$ of homogeneous polynomial functions, i.e.  for each $\rho=1,\dots,r$ we have $F^{\rho}\in\mathcal{P}_{k_{\rho}}(\pi)$ for some $k_{\rho} \in \Bbb Z$. Then the ideal 
$$I(\mathcal{E})=\left\{\psi\in\mathcal{P}(\pi)\ |\ \psi=\sum_{\rho,\sigma}\psi_{\rho,\sigma}{D}_{\sigma}(F^{\rho}),\psi_{\rho,\sigma}\in\mathcal{P}(\pi)\right\},
$$ 
(cf. \cite{boch} for more details), is a homogeneous ideal, thus the quotient $\mathbb{R}$-algebra $\mathcal{P}(\mathcal{E})=\mathcal{P}(\pi)/I(\mathcal{E})$ of all polynomial functions on $J^{\infty}(\pi)$ restricted to $\mathcal{E}$ inherits the $\mathbb{Z}$-grading from $\mathcal{P}(\pi)$, the weight of a coset $\psi+I(\mathcal{E})$ being $|\psi+I(\mathcal{E})|=|\psi|$.

Repeating all the constructions made above for the $\Bbb Z$-graded ring $\mathcal{P}(\mathcal{E})$, the $\mathcal{P}(\mathcal{E})$-module structure and the Lie algebra structure on the set $\mathrm{Vect}_{\mathcal{P}(\mathcal{E})}(\pi)$ of all vector fields restricted to $\mathcal{E}$ as well as the $\mathcal{P}(\mathcal{E})$-module structure and the Lie algebra  structure on the set $\mathcal{P}^m(\mathcal{E})$ of all m-tuples of polynomial functions (generating functions) restricted to $\mathcal{E}$ inherit the $\mathbb{Z}$-grading too. 

Hence, \textit{a generating function} $\bar\Phi=(\phi_1+I(\mathcal{E}),\dots,\phi_m+I(\mathcal{E}))\in \mathcal{P}^m(\mathcal{E})$ of an evolutionary vector field is \textit{homogeneous of the weight} $k$ if and only if all its nonzero components satisfy the conditions \eqref{weight_evol_field}.

\subsection{The $\mathbb{Z}$-grading on the Lie algebra ${\mathrm{sym}_{\mathcal{P}(\mathcal{E})}(\mathcal{E}})$ of generating functions of symmetries of an equation $\mathcal{E}$}
Let us consider an equation $\mathcal{E}=\mathcal{E}_F$ defined by an $r$-tuple $F=(F^1,\dots,F^r)$ of homogeneous polynomial functions. We claim that 
$$\mathrm{sym}_{\mathcal{P}(\mathcal{E})}(\mathcal{E})=\bigoplus_{k\in\mathbb{Z}}\mathrm{sym}_{\mathcal{P}(\mathcal{E}),k}(\mathcal{E}),$$ 
where $\mathrm{sym}_{\mathcal{P}(\mathcal{E}),k}(\mathcal{E})=\left\{\bar\Phi\in\mathcal{P}^m_k(\mathcal{E})\ |\ \bar\ell_{F}(\bar\Phi)=0\right\}$ is the vector space of all polynomial generating functions $\bar\Phi$ of symmetries of $\mathcal{E}$ that are (considered as $m$-tuples of polynomials) homogeneous of the weight $k$. 

To see this, firstly note that the \textit{linearization operator} $\ell_{F}:\mathcal{P}^m(\pi)\to\mathcal{P}^r(\pi)$ acts as follows: the image of a homogeneous $m$-tuple $\Phi=(\varphi_1,\dots,\varphi_m)\in\mathcal{P}^m_k(\pi)$, $k \in \Bbb Z$, is an $r$-tuple $\ell_{F}(\Phi)=(\psi_1,\dots,\psi_{r})\in\mathcal{P}^r(\pi)$ of homogeneous polynomials, their weights being \begin{equation}\label{weight_lin}|\psi_{\rho}|=|F^\rho|+k \end{equation} for all $\rho=1,\dots,r.$ The restriction $\bar\ell_{F}:\mathcal{P}^m(\mathcal{E})\to\mathcal{P}^r(\pi)$ evinces the same behavior.  

Further, let $\bar\Phi\in\mathrm{sym}_{\mathcal{P}(\mathcal{E})}(\mathcal{E})$ be a symmetry, i.e.  $\bar\ell_F(\bar\Phi)=0$. Since $\mathrm{sym}_{\mathcal{P}(\mathcal{E})}(\mathcal{E})\subset\mathcal{P}^m(\mathcal{E})=\bigoplus_{k\in\mathbb{Z}}\mathcal{P}^m_k(\mathcal{E})$, we can unambiguously decompose the $m$-tuple $\bar \Phi$ as a sum of homogeneous $m$-tuples of different weights, that is $\bar\Phi=\bar\Phi^1+\dots+\bar\Phi^l$, where $|\bar\Phi^{\eta}|\neq|\bar\Phi^{\epsilon}|$ for $\eta\neq\epsilon$. 
The aim is to show that each of these homogeneous $m$-tuples is a symmetry of $\mathcal{E}$.

Nevertheless, from the linearity of $\bar\ell_F$ it holds $\bar\ell_F(\bar\Phi)=\bar\ell_F(\bar\Phi^1)+\dots+\bar\ell_F(\bar\Phi^l) = 0$, thus we have $(\bar{\ell}_F(\bar\Phi))_{\rho} =(\bar{\ell}_F(\bar\Phi^{1}))_{\rho}+\dots+(\bar\ell_F(\bar\Phi^{l}))_{\rho} = 0$ for each $\rho$-th component of $\bar\ell_F(\bar\Phi)$, $\rho=1,\ldots,r$. Moreover, it follows from \eqref{weight_lin} that $|(\bar\ell_F(\bar\Phi^{\eta}))_{\rho}|=|F^{\rho}|+|\bar\Phi^{\eta}|,$ which together with the fact that $|\bar\Phi^{\eta}|\neq|\bar\Phi^{\epsilon}|$ for $\eta\neq\epsilon$ gives us $|(\bar\ell_F(\bar\Phi^{\eta}))_{\rho}|\neq |(\bar\ell_F(\bar\Phi^{\epsilon}))_{\rho}|$ for $\eta\neq\epsilon$. Thus, $(\bar\ell_F(\bar\Phi^{\eta}))_{\rho}$ takes the form of a sum of the homogeneous polynomials $(\bar{\ell}_F(\bar\Phi^{j}))_{\rho}$, $j=1,\ldots, l$, of different weights, and this sum vanishes. But this is possible if and only if each summand $(\bar{\ell}_F(\bar\Phi^{j}))_{\rho}$ is equal to zero, which means that $\bar{\ell}_F(\bar\Phi^{j})=((\bar{\ell}_F(\bar\Phi^{j}))_{1}, \dots,(\bar{\ell}_F(\bar\Phi^{j}))_{r})=0$. Hence, we conclude that $\bar\Phi^j$ is a symmetry for all $j=1,\dots,l$.

\begin{remark}
Note that all the constructions on $\mathcal{E}$ are done in the external coordinates, i.e. in the coordinates on $J^{\infty}(\pi)$. This topic in view of the internal coordinates will be discussed separately in Sect. \ref{propRO}.
\end{remark}

\section{Synopsis: Covering and recursion operators for full-fledged nonlocal symmetries of the rYME}
\label{covrYME}
Starting from this section, we turn our attention to the rYME \eqref{yme}. The main aim is to continue our previous research of the equation in question, see \cite{jahn_voj} for more details, and to demonstrate the usefulness of the full-fledged recursion operators in the study of the Lie algebras of the full-fledged nonlocal symmetries. First of all, we very briefly summarize the main results of \cite{jahn_voj} which are frequently used in the following sections.

From now on, let $\mathcal{E}$ denote the rYME \eqref{yme} together with all its differential consequences. So, $\mathcal{E}$ is a submanifold in the jet space $J^{\infty}(\pi)$ over the bundle $\pi:\mathbb{R}^5\to\mathbb{R}^4$,  where $x,y,z,t$ denote the independent variables and $u$ stands for the dependent variable. The internal coordinates on $\mathcal{E}$ are chosen to be $u_{x^it^j},u_{x^iy^kt^j},u_{x^iz^lt^j}$, where $i,j\geq 0$ and  $k,l>0$.
Expanding the two-parameter Lax pair \cite{jahn_voj}, one can obtain the infinite-dimensional differential covering $\tau^W:\tilde{\mathcal{E}}^W\to\mathcal{E}$ of $\mathcal{E}$, where $\tilde{\mathcal{E}}^W\subset J^{\infty}(\tilde\pi)$ is given by the following system of differential equations (again considered including all their differential consequences):
\begin{equation}
\begin{aligned}
\label{rYMEcov}
\tilde{\mathcal{E}}^W:
		u_{yz} &= u_{tx} - u_z u_{xx}+u_xu_{xz},\\
              q_{\alpha,t}&= u_zq_{\alpha,x} + q_{\alpha-1,z},\\[2pt]
                q_{\alpha,y} &= u_xq_{\alpha,x} + q_{\alpha-1,x},\\
              m_{\alpha,t} &= u_z m_{\alpha,x}-\displaystyle \frac{z}{t}m_{\alpha,z}-\frac{1}{t}m_{\alpha-1,z},\\[2mm]
                 m_{\alpha,y} &= u_x m_{\alpha,x}-\displaystyle \frac{z}{t}m_{\alpha,x}-\frac{1}{t}m_{\alpha-1,x},\\[2pt]
                 r_{\beta,x} &= u_xr_{\beta-1,x} - r_{\beta-1,y},\\
               r_{\beta,z} &= u_zr_{\beta-1,x} - r_{\beta-1,t}, \qquad  \qquad \alpha \geq 0,\ \beta \geq 1.
 \end{aligned}
 \end{equation}
Here $q_{-1} := 0,\   m_{-1}: = 0,\  r_{-1} := x,\  r_0:=-u$, and $q_{\alpha},\ m_{\alpha},\ r_{\beta}$ are additional dependent variables that give rise to nonlocal variables. The internal coordinates on $\tilde{\mathcal{E}}^W \subset J^{\infty}(\tilde\pi)$ {are} $u_{x^it^j},u_{x^iy^kt^j},u_{x^iz^lt^j}, q_{\alpha,x^iz^j}, m_{\alpha,x^iz^j}, r_{\beta,y^it^j}$, where $i,j\geq 0$ and $k,l>0$. The $\mathbb{R}$-algebra of all functions on $\tilde{\mathcal{E}}^W$ that are locally smooth as the functions of the internal coordinates (explicitly depending on just finitely many of them) on $\tilde{\mathcal{E}}^W$ will be denoted as $\mathcal{A}$.

The full-fledged nonlocal $\tau^W$-symmetries (hereafter we will call them shortly $\tau^W$-symme\-tries) of the rYME \eqref{yme} are the vector fields
$$\mathbf{E}_P=\mathbf{E}^W_{p_0}+\sum_{\alpha=0}^{\infty}\sum_{a,b=0}^{\infty}\left(\tilde D_x^a\tilde D_z^b(p_{\alpha}^{q})\frac{\partial}{\partial q_{\alpha,x^az^b}}+\tilde D_x^a\tilde D_z^b(p_{\alpha}^{m})\frac{\partial}{\partial m_{\alpha,x^az^b}}+\tilde D_y^a\tilde D_t^b(p_{\alpha+1}^{r})\frac{\partial}{\partial r_{\alpha+1,y^at^b}}\right),$$
where
$$\mathbf{E}^W_{p_0}=\sum_{
{\substack{i,j\geq 0,\\[1mm] k,l>0}}}\left(\tilde D_x^i\tilde D_z(p_0)\frac{\partial}{u_{x^it^j}}+\tilde D_x^i\tilde D_y^k\tilde D_z^j(p_0)\frac{\partial}{\partial u_{x^iy^kt^j}}+\tilde D_x^i\tilde D_z^l\tilde D_t^j(p_0)\frac{\partial}{\partial u_{x^iz^lt^j}}\right),$$
and  $p_0, p_{\alpha}^{q},p_{\alpha}^{m}, p_1^{r}, p_{\beta}^{r}$, $\alpha\geq 0,\beta\geq 2$, are functions from $\mathcal{A}$ satisfying the conditions\\
\begin{align}
\label{nsym-yme1}
\tilde D_{yz}(p_0) - \tilde D_{tx}(p_0) + u_z \tilde D_{xx}(p_0)-u_x\tilde D_{xz}(p_0)+u_{xx}\tilde D_z(p_0)-u_{xz}\tilde D_x(p_0)&=0,\tag{NS1}\\[2mm]
\label{nsym-yme2}
\tilde D_t(p_\alpha^q) - u_z\tilde D_x(p_\alpha^q) - q_{\alpha,x}\tilde D_z(p_0)-\tilde D_z(p_{\alpha-1}^q)&=0\tag{NS2},\\[2pt]
\label{nsym-yme3}
\tilde D_y(p_\alpha^q) -u_x\tilde D_x(p_\alpha^q) - q_{\alpha,x}\tilde D_x(p_0)-\tilde D_x(p_{\alpha-1}^q)&=0\tag{NS3},\\[2pt]
\label{nsym-yme4}
\tilde D_t(p_\alpha^m) -u_z\tilde D_x(p_\alpha^m) - m_{\alpha,x}\tilde D_z(p_0)+\frac{z}{t}\tilde D_z(p_\alpha^m)+\frac{1}{t}\tilde D_z(p_{\alpha-1}^m)&=0\tag{NS4},\\[2mm]
\label{nsym-yme5}
\tilde D_y(p_\alpha^m) -u_x\tilde D_x(p_\alpha^m) - m_{\alpha,x}\tilde D_x(p_0)+\frac{z}{t}\tilde D_x(p_\alpha^m)+\frac{1}{t}\tilde D_x(p_{\alpha-1}^m)&=0\tag{NS5},\\[2mm]
\label{nsym-yme6}
\tilde D_x(p_1^r) - \tilde D_y(p_0)+2u_x\tilde D_x(p_0)&=0\tag{NS6},\\[2pt]
\label{nsym-yme7}
\tilde D_z(p_1^r) -\tilde D_t(p_0)+u_x\tilde D_z(p_0)+u_z\tilde D_x(p_0)&=0\tag{NS7},\\[2pt]
\label{nsym-yme8}
\tilde D_x(p_\beta^r) - u_x\tilde D_x(p_{\beta-1}^r)-r_{\beta-1,x}\tilde D_x(p_0)+D_y(p_{\beta-1}^r)&=0\tag{NS8},\\[2pt]
\label{nsym-yme9}
\tilde D_z(p_\beta^r) - u_z\tilde D_x(p_{\beta-1}^r)-r_{\beta-1,x}\tilde D_z(p_0)+D_t(p_{\beta-1}^r)&=0,\tag{NS9}
\end{align}
where $p_{-1}^q=p_{-1}^m=0$.

Any $\tau^W$-symmetry is (uniquely) represented by the vector-valued generating function 
$$P=\left[p_0, p_0^q, p_0^m, p_1^r,p_1^q, p_1^m, p_2^r,p_2^q, p_2^m\ldots  \right]\in\mathcal{A}^{\mathbb{N}},$$
thus, as usual, we do not distinguish between symmetries and their generating functions below. Hence, we can consider the Lie algebra $\mathrm{sym}^{\tau^W}_{\mathcal{A}}(\mathcal{E})$ of the $\tau^W$-symmetries of $\mathcal{E}$ (with the generating functions from $\mathcal{A}^{\mathbb{N}}$) to be the Lie subalgebra of the Lie algebra $\mathcal{A}^{\mathbb{N}}$ endowed with the Jacobi bracket \eqref{sec1:2}, i.e. $\mathrm{sym}^{\tau^W}_{\mathcal{A}}(\mathcal{E})=\mathrm{ker}\ \ell_{\tilde{\mathcal{E}}^W}\subset\mathcal{A}^{\mathbb{N}}.$

In \cite{jahn_voj}, we introduced two  automorphisms of the $\mathcal{A}$-module $\mathcal{A}^{\mathbb{N}}$, namely the mappings $\mathcal{R}^q, \mathcal{R}^m:\mathcal{A}^{\mathbb{N}}\to\mathcal{A}^{\mathbb{N}}$, $P\mapsto \hat P$, given by the formulas
\begin{align}
\label{ROq}
 \mathcal{R}^q:&\ 
                \left\{\begin{array}{l}
     		\hat p_0 = u_x p_0+p_1^r,\\[1.5mm]
		\hat p_\alpha^q = q_{\alpha,x} p_0 + p_{\alpha-1}^q,\\[1.5mm]
		\hat p_\alpha^m = m_{\alpha,x} p_0 -\frac{1}{t} p_{\alpha-1}^m - \frac{z}{t}  p_\alpha^m ,\\[1.5mm]
		\hat p_\beta^r = r_{\beta,x} p_0 - p_{\beta+1}^r,
                \end{array}\right.
\intertext{and}
\label{ROm}
 \mathcal{R}^m&=t\mathcal{R}^q+z\mathcal{I},
\end{align}
where $\alpha\in\mathbb{N}_0$, $\beta \in\mathbb{N}$, $\hat p_0^r=-\hat p_0$, and $\mathcal{I}$ denotes the identity mapping on $\mathcal{A}^{\Bbb N}$, their  inverses reading as follows:
\begin{align}
\label{InvROq}
(\mathcal{R}^q)^{-1}:&\ 
                \left\{\begin{array}{l}
     		p_0 = \frac{\hat p_0^q}{q_{0,x}},\\[2mm]
		p_\alpha^q = -\frac{q_{\alpha+1,x}}{q_{0,x}} \hat p_0^q + \hat p_{\alpha+1}^q,\\[2mm]
		p_\alpha^m =  \sum _{j=0}^\alpha \frac{t\hat p_j^m}{(-z)^{\alpha-j+1}}-\frac{t \hat p_0^q}{q_{0,x}} \sum _{j=0}^\alpha \frac{m_{j,x}}{(-z)^{\alpha-j+1}},\\[2mm]
		p_\beta^r = \frac{r_{\beta-1,x}}{q_{0,x}}\hat p_0^q - \hat p_{\beta-1}^r, \quad \alpha \geq 0;
                \end{array}\right.
\intertext{and}
\label{InvROm}
(\mathcal{R}^m)^{-1}:&\ 
                \left\{\begin{array}{l}
     		p_0 = \frac{\hat p_0^m}{tm_{0,x}},\\[2mm]
		p_\alpha^q = \sum_{j=0}^{\alpha}\frac{(-t)^{{\alpha-j}}\hat p_j^q}{z^{\alpha-j+1}}- \frac{\hat p_{0}^m}{m_{0,x}}\sum_{j=0}^{\alpha}\frac{(-t)^{\alpha-j}q_{j,x}}{z^{\alpha-j+1}},\\[2mm]		
		p_\alpha^m =\frac{m_{\alpha+1,x}}{m_{0,x}}\hat p_0^m-\hat p_{\alpha+1}^m,\\[2mm]		
		p_\beta^r = -\sum_{j=0}^{\beta-1}\frac{z^{\beta-j-1}\hat p_j^r}{t^{\beta-j}}+\frac{\hat p_0^m}{m_{0,x}}\left(-\frac{z^{\beta}}{t^{\beta+1}}+\sum_{j=1}^{\beta}\frac{z^{\beta-j}r_{j-1,x}}{t^{\beta-j+1}}\right).
                \end{array}\right.
\end{align}

It was proved that all the ($\mathbb{R}$-linear) maps $\mathcal{R}^q$, $\mathcal{R}^m$, $(\mathcal{R}^q)^{-1}$ and $(\mathcal{R}^m)^{-1}$ leave the set ${\mathrm{sym}}^{\tau^W}_{\mathcal{A}}(\mathcal{E})$ invariant, hence, they are the recursion operators for $\tau^W$-symmetries of $\mathcal{E}$. Since an arbitrary composition of these operators provides us with another recursion operator, we have introduced the following notation:
$$\mathcal R_i^j = \underbrace{({\mathcal R}^q)^{-1} \circ \ldots \circ ({\mathcal R}^q)^{-1}}_{i-times} \circ \underbrace{\mathcal R^m \circ \ldots \circ \mathcal R^m}_{j-times}, \qquad i,j \in \Bbb Z,$$
where for negative values of $i$, resp. $j$, we define
$$\underbrace{({\mathcal R}^q)^{-1} \circ \ldots \circ ({\mathcal R}^q)^{-1}}_{i-times}=\underbrace{\mathcal R^q \circ \ldots \circ \mathcal R^q}_{-i-times}, \quad \mathrm{resp.}\ \underbrace{\mathcal R^m \circ \ldots \circ \mathcal R^m}_{j-times}=\underbrace{({\mathcal R}^m)^{-1} \circ \ldots \circ ({\mathcal R}^m)^{-1}}_{-j-times}.$$ 
 
It was shown in \cite{jahn_voj} that the set $\{\mathcal{R}_i^j\}_{i,j\in\mathbb{Z}}$ equipped with the composition operation $\circ$ forms an abelian group, that is
$$\mathcal R_i^j \circ \mathcal R_k^l = \mathcal R_k^l \circ \mathcal R_i^j = \mathcal R_{i+k}^{j+l},$$ 
the inverse being $(\mathcal R_i^j)^{-1}=\mathcal R_{-i}^{-j}$. To refer to this group, we will use the standard notation for the group generated by elements $\mathcal{R}^q$ and  $\mathcal{R}^m$, i.e. $\langle\mathcal{R}^q,\mathcal{R}^m\rangle:=(\{\mathcal{R}_i^j\}_{i,j\in\mathbb{Z}},\circ)$. Similarly,  given fixed integers $i,j\in\mathbb{Z}$, the cyclic subgroup $\{(\mathcal{R}_i^j)^k\ |\ k\in\mathbb{Z}\}=\{\mathcal{R}_{k\cdot i}^{k\cdot j}\ |\ {k\in\mathbb{Z}}\}$ of the group $\langle\mathcal{R}^q,\mathcal{R}^m\rangle$ generated by the element $\mathcal{R}_i^j$ will be denoted as $\langle\mathcal{R}_i^j\rangle$.

\begin{remark}
In the two-indices notation of the recursion operators, we obviously have the following equivalences:
$$\mathcal R^q \equiv \mathcal R_{-1}^0, \qquad ({\mathcal R}^q)^{-1} \equiv \mathcal R_1^0, \qquad \mathcal R^m \equiv \mathcal R_0^1, \qquad ({\mathcal R}^m)^{-1} \equiv \mathcal R_0^{-1},\qquad \mathcal I \equiv \mathcal R_0^0.$$
Below, we will refer to these recursion operators as the \textit{basic recursion operators}. For each of them, we will use both equivalent notations in turn, according to the situation.
 \end{remark}

\section{The properties of the recursion operators $\mathcal{R}_i^j$ restricted to the $\mathcal{L}$-module $\mathcal{L}^{\mathbb{N}}$ of sequences of Laurent  polynomial functions on $\tilde{\mathcal{E}}^W$.}
\label{propRO}
In order to gain more information about the action of the operators $\mathcal{R}_i^j$ on the $\mathcal{A}$-module $\mathcal{A}^{\mathbb{N}}$, and hence also on the Lie algebra $\mathrm{sym}_{\mathcal{A}}^{\tau^W}(\mathcal{E})\subset\mathcal{A}^{\mathbb{N}}$, we will focus on its submodule that can be equipped with a $\mathbb{Z}$-grading. As we are working in the internal coordinates on $\tilde{\mathcal{E}}^{W}$ and $\mathcal{A}$ is the $\mathbb{R}$-algebra of functions that are locally smooth as functions of the internal coordinates, it is natural to focus on the set of sequences of Laurent polynomial functions.
Let $\mathcal{L}$ denote the ring of those functions on $J^{\infty}(\tilde\pi)$ that are Laurent polynomials in the internal variables $x,y,z,t,u_{x^it^j},u_{x^iy^kt^j},u_{x^iz^lt^j}, q_{\alpha,x^az^b},m_{\alpha,x^az^b},r_{\beta,y^at^b}$, where $i,j,a,b,\alpha \geq 0$, $k,l,\beta>0$. The object of our interest is therefore the $\mathcal{L}$-module $\mathcal{L}^{\mathbb{N}}$.

The ring $\mathcal{L}$ and the $\mathcal{L}$-module  $\mathcal{L}^{\mathbb{N}}$ can be endowed with the $\mathbb{Z}$-grading exactly in the same way as the ring $\mathcal{L}(\pi)$ and the $\mathcal{L}(\pi)$-module $\mathcal{L}^m(\pi)$  in the exposition of Sect. \ref{grading} (allowing $m$ to be infinite). 
However, recall that the primary focus of our study is on  $\tau^W$-symmetries  whose generating functions lie in $\mathcal{L}^{\mathbb{N}}$. Note that many of these symmetries arise from $\tau^W$-symmetries that are polynomial in the external coordinates. Indeed, consider for a moment the jet space $J^{\infty}(\tilde\pi)$ and its coordinates. Let $\mathbf{E}_P$ be a $\tau^W$-symmetry of $\mathcal{E}$ with the generating function $P\in\mathcal{P}^{\mathbb{N}}({\tilde{\mathcal{E}}}^{W})$, where $\mathcal{P}^{\mathbb{N}}({\tilde{\mathcal{E}}}^{W})$ is the Lie algebra of sequences of polynomial functions in these external coordinates, restricted to ${\tilde{\mathcal{E}}}^{W}$. Then $\mathbf{E}_P$, when expressed in the internal coordinates on  $\tilde{\mathcal{E}}^{W}$, has a generating function that belongs to the space $\mathcal{L}^{\mathbb{N}}$. Therefore, it is natural to require the gradings on $\mathcal{L}^{\mathbb{N}}$  and $\mathcal{P}^{\mathbb{N}}({\tilde{\mathcal{E}}}^{W})$ to be coherent, meaning that the 
symmetry $\mathbf{E}_P\in\mathrm{sym}_{\mathcal{P}({\tilde{\mathcal{E}}}^{W})}({\tilde{\mathcal{E}}}^{W})$ is homogeneous if and only if $\mathbf{E}_{P,\mathrm{int}}$ expressed in the internal coordinates on ${\tilde{\mathcal{E}}}^{W}$ is homogeneous, and their weights coincide, i.e., $|P|=|\mathbf{E}_P|=|\mathbf{E}_{P,\mathrm{int}}|=|P_{\mathrm{int}}|$.

To this end, it is necessary to choose the weight setting \eqref{weight} such that all the functions defining the system \eqref{rYMEcov} are homogeneous. Indeed, having chosen such a weight setting, we are provided with both the $\mathbb{Z}$-grading on $\mathcal{L}$ and the $\mathbb{Z}$-grading on the ring $\mathcal{P}({\tilde{\mathcal{E}}}^{W})$.  Let $\phi_{\mathrm{int}}$ denote a function $\bar\phi=\phi+I({\tilde{\mathcal{E}}}^{W})$ on ${\tilde{\mathcal{E}}}^{W}$ expressed in the internal coordinates. Then we can see that 
$\bar\phi=\phi+I({\tilde{\mathcal{E}}}^{W})\in\mathcal{P}({\tilde{\mathcal{E}}}^{W})$  is homogeneous if and only if the function $\phi_{\mathrm{int}}\in\mathcal{L}$  is homogeneous, and $|\bar\phi|=|\phi_{\mathrm{int}}|$. Thus, \textit{inter alia}, the weight $|P|$ of a polynomial symmetry $P\in\mathrm{sym}_{\mathcal{P}({\tilde{\mathcal{E}}}^{W})}({\tilde{\mathcal{E}}}^{W})$ and the weight $|P_{\mathrm{int}}|$ of the symmetry $P$ expressed in the internal coordinates coincide, as we require.

Moreover, analogously as in Sect. \ref{grading}, we can immediately deduce that the linearization operator $\bar\ell:\mathcal{P}^{\mathbb{N}}({\tilde{\mathcal{E}}}^{W})\to  {\mathcal{P}}^{\mathbb{N}}(\tilde\pi)$ as well as the linearization operator $\bar\ell_{\mathrm{int}}:\mathcal{L}^{\mathbb{N}}\to\mathcal{L}^{\mathbb{N}}$ expressed in the internal coordinates maps homogeneous sequences to sequences of homogeneous functions. 
In accordance with the notation used above, let $\mathrm{sym}_{\mathcal{L}}^{\tau^W}(\mathcal{E})=\{P\in\mathcal{L}^{\mathbb{N}}\ |\ \bar\ell_{\mathrm{int}}(P)=0\}$ denote the Lie algebra of symmetries whose generating functions are the sequences of differential Laurent polynomial functions of the internal coordinates on ${\tilde{\mathcal{E}}}^{W}$,  and let $\mathrm{sym}_{\mathcal{L},k}^{\tau^W}(\mathcal{E})$ denote the set of symmetries $P\in\mathrm{sym}_{\mathcal{L}}^{\tau^W}(\mathcal{E})$ that are homogeneous of the weight $k$.
 Similarly as in Sect. \ref{grading} we can conclude that $\mathrm{sym}_{\mathcal{L}}^{\tau^W}(\mathcal{E})=\bigoplus_{k\in\mathbb{Z}}\mathrm{sym}_{\mathcal{L},k}^{\tau^W}(\mathcal{E})$.

Thus, from now on, we  focus just on the $\mathcal{L}$-module $\mathcal{L}^{\mathbb{N}}\subset{\mathcal{A}}^{\mathbb{N}}$ and on the action of the recursion operators $\mathcal{R}^i_j$ on it. To make $\mathcal{L}^{\mathbb{N}}$ $\mathbb{Z}$-graded, we have to choose an appropriate weight setting \eqref{weight}.
The requirement for the functions defining our system \eqref{rYMEcov} to be homogeneous yields the following system of independent conditions:
\begin{equation}
\label{weight-cond}
\begin{aligned}
&|x|+|z|-|t|-|u|=0,\\
&|y|+2|z|-2|t|-|u|=0,\\
&|q_{\alpha}|-|q_{\alpha-1}|+|z|-|t|=0,\\
&|m_{\alpha}|-|m_{\alpha-1}|+|z|=0,\\
&|r_{\beta}|-|r_{\beta-1}|-|z|+|t|=0,\quad \alpha\geq 0,\, \beta\geq 1.
\end{aligned}
\end{equation}

It turns out that, among all possible solutions of the system \eqref{weight-cond}, the following choice of weights is suitable for our purposes: 
\begin{equation}
\label{weightchoice}
|t|=|x|=1,\quad |y|=2,\quad |z|=|u|=|m_{\alpha}|=0,\quad |q_{\alpha}|=\alpha,\quad  |r_{\beta}|=-\beta.
\end{equation}

Keeping in the mind the discussion concerning the internal and external variables made above, and following the exposition made in Sect. \ref{grading}, the weight setting \eqref{weightchoice} provides us with the $\mathbb{Z}$-grading on the $\mathcal{L}$-module $ \mathcal{L}^{\mathbb{N}}$. 
Therefore, we have $\mathcal{L}^{\mathbb{N}}=\bigoplus_{k\in\mathbb{Z}}{\mathcal{L}^{\mathbb{N}}_k}$, where ${\mathcal{L}_k^{\Bbb N}}$ denotes the vector space of all homogeneous sequences of differential Laurent polynomials of the internal variables on $\tilde{\mathcal{E}}^{W}$ of the weight $k$. 
Hence, upon the weight setting \eqref{weightchoice}, the formula \eqref{weight_evol_field} gives us that 
$P=[p_0,p_0^q,p_0^m,p_1^r,p_1^q,p_1^m,p_2^r,p_2^q,p_2^m,\dots] \in \mathcal{L}^{\mathbb{N}}$
is homogeneous of the weight $|P|=k$ if and only if all its nonzero components are homogeneous, and their weights satisfy the conditions
\begin{equation}
\label{weights_of_P}
|p_0|=|p_{\alpha}^m|=k,\quad |p_\alpha^q|=k+\alpha,\quad |p_\beta^r|=k-\beta. \\
\end{equation}

The following proposition describes the action of the basic recursion operators $\mathcal{R}^q$ and $\mathcal{R}^m$ on the $\mathbb{Z}$-graded $\mathcal{L}$-module $\mathcal{L}^{\mathbb{N}}$.
\begin{proposition}
\label{grad01} 
Consider the $\mathbb{Z}$-graded $\mathcal{L}$-module
$\mathcal{L}^{\mathbb{N}}=\bigoplus_{k\in\mathbb{Z}}\mathcal{L}_k^{\mathbb{N}}$
 determined by the weight setting \eqref{weightchoice}. Let $P = [p_{0},p_{0}^q,p_{0}^m,p_{1}^r,p_{1}^q,p_{1}^m,p_{2}^r,p_{2}^q,p_{2}^m,\ldots] 
\in\mathcal{L}_k^{\mathbb{N}}$
 be a homogeneous sequence. Then:
\begin{itemize}
\item[a)] $\mathcal{R}^q(P) \in  \mathcal{L}_{k-1}^{\mathbb{N}}$, i.e. the mapping $\mathcal{R}^q: {\mathcal{L}}^{\mathbb{N}}\to{\mathcal{L}}^{\mathbb{N}}$ is a graded $\mathcal{L}$-module automorphism of degree $-1$;
\item[b)] $\mathcal{R}^m(P) \in \mathcal{L}_k^{\mathbb{N}}$, i.e. the mapping $\mathcal{R}^m: {\mathcal{L}}^{\mathbb{N}}\to{\mathcal{L}}^{\mathbb{N}}$ is a graded $\mathcal{L}$-module automorphism of degree $0$.
\end{itemize}
\end{proposition}

\begin{proof}
a)  Let $P$ be homogeneous with the weight $|P|=k$, which implies that the weights of all its nonzero components are well defined and satisfy the conditions \eqref{weights_of_P}. Let $\hat P = [\hat p_{0},\hat p_{0}^q,\hat p_{0}^m,\hat p_{1}^r,p_{1}^q,\hat p_{1}^m,\hat p_{2}^r,\hat p_{2}^q,\hat p_{2}^m,\dots]= \mathcal{R}^q(P)$. Our aim is to show that all nontrivial components of $\hat P$ are homogeneous and their weights satisfy the conditions \eqref{weights_of_P}  with $k$ replaced with $k-1$.

 If $\hat p_0=u_xp_0+p_1^r$ is nontrivial then we have either $p_{0} \neq 0$ or $p_{1}^r \neq 0$. Hence, whenever the terms below are non-zero (and thus their weights are well-defined), we can calculate
\begin{align*}
|u_x p_{0}| &= |p_{0}| + |u| - |x| = k-1,
\intertext{resp.}
|p_{1}^r| &=  k-1,
\end{align*} 
so both the summands $u_xp_0$ and $p_1^r$, if nontrivial, have the same weight. Thus, $\hat p_0$ is homogeneous with the weight $|\hat p_0|=k-1$.

Further, let $\alpha \in \Bbb N_0$, and let $\hat p_\alpha^q=q_{\alpha,x}p_0+p_{\alpha-1}^q$ be nontrivial. Then we have either $p_{0} \neq 0$ or $p_{\alpha-1}^q \neq 0$, and similarly as above we obtain 
\begin{align*}
|q_{\alpha,x} p_{0}| &= |p_{0}|+ |q_\alpha| - |x| = k+\alpha-1,
\intertext{resp.}
|p_{\alpha-1}^q| &=  k+\alpha-1.
\end{align*} 
This means that both the summands $q_{\alpha,x}p_0$ and $p_{\alpha-1}^q$,  if nontrivial, have the same weight. Thus, $\hat p_\alpha^q$ is homogeneous with the weight $|\hat p_\alpha^q|= k+\alpha-1$.

Similarly, using the formulas \eqref{ROq} and \eqref{weights_of_P} one obtains that $\hat p_\alpha^m$ is homogeneous with the weight $|\hat p_\alpha^m|=k-1$ and $\hat p_\beta^r$ is homogeneous with the weight $|\hat p_\beta^r|=k-\beta+1$. Hence, we have
$$|\hat p_0| = |\hat p_\alpha^q|-\alpha = |\hat p_\alpha^m|= |\hat p_\beta^r|+\beta = k-1,$$
and, according to \eqref{weights_of_P}, the latter means that $\hat P$ is homogeneous of the weight $|\hat P|=k-1$.\\ 

b) Let $|P|=k$. Then, using the result of the paragraph a), we can calculate (recall that $\mathcal R^m = t\mathcal R^q+z\mathcal I$, where $\mathcal I$ denotes the identity operator)
\begin{align*}
|t\mathcal R^q(P)| &= |t| + |\mathcal R^q(P)| = 1+k-1=k,
\intertext{and}
|z\mathcal I(P)| &= |z| + |\mathcal I(P)| = 0+k=k.
\end{align*} 
This immediately implies that $\mathcal R^m(P)$ is homogeneous and $|\mathcal R^m(P)|=k$ .
\end{proof}

\begin{corollary}
\label{corgrad01}
Given the weight setting \eqref{weightchoice}, the inverse mappings $(\mathcal{R}^q)^{-1}$, resp. $(\mathcal{R}^m)^{-1}$, are graded $\mathcal{L}$-module automorphisms of degrees $1$, resp. $0$.
\end{corollary}
\begin{corollary}
\label{degRij}
Given the weight setting \eqref{weightchoice}, the mapping $\mathcal{R}_i^j$, where $i,j \in \Bbb Z$, is a graded $\mathcal{L}$-module automorphism of degree $i$.
\end{corollary}

\begin{figure}[t]
\begin{center}
\begin{tikzcd}[row sep=2.0cm, column sep=2.5cm]
\mathcal L^{\Bbb N}_k
\arrow[bend left=15, "\mathcal R_1^0" description]{r} 
\arrow[bend right=20, "\mathcal R_0^{-1}" description]{d} 
&
\mathcal L^{\Bbb N}_{k+1}
\arrow[bend left=15, "\mathcal R_{-1}^0" description]{l} 
\arrow[bend right=20, "\mathcal R_0^{-1}" description]{d}
\\ 
\mathcal L^{\Bbb N}_k
\arrow[bend left=15, "\mathcal R_1^0" description]{r} 
\arrow[bend right=20, "\mathcal R_0^1" description]{u}
&
\mathcal L^{\Bbb N}_{k+1}
\arrow[bend left=15, "\mathcal R_{-1}^0" description]{l} 
\arrow[bend right=20, "\mathcal R_0^1" description]{u}
\end{tikzcd}
\end{center}
  \caption{Change of weights under the action of the basic recursion operators}
  \label{fig:1}
\end{figure}
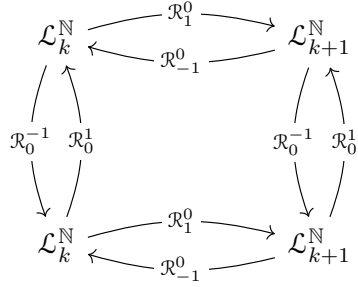

The results stated in  Prop. \ref{grad01} and Cor. \ref{corgrad01} can be schematically summarized by means of a diagram, see Fig. \ref{fig:1}. As all the operators $\mathcal{R}_i^j$, $i,j \in \Bbb Z$, mutually commute (see Sect. \ref{covrYME}), this diagram is commutative.

\begin{remark}
In what follows,
given any subgroup $G$ of the group $\langle\mathcal{R}^q,\mathcal{R}^m\rangle$, the orbit of an $\tau^W$-symmetry $P$ under the action of $G$ on the set $\mathrm{sym}_{\mathcal{A}}^{\tau^W}(\mathcal E)$ will be denoted as $G\cdot P$. Thus, in particular, $\langle\mathcal{R}_i^j\rangle\cdot P$ denotes the orbit of the element $P\in\mathrm{sym}_{\mathcal{A}}^{\tau^W}(\mathcal E)$ under the action of $\langle\mathcal{R}_i^j\rangle$, i.e. $\langle\mathcal{R}_i^j\rangle\cdot P=\{\mathcal{R}_{k\cdot i}^{k\cdot j}(P)\ | \ k \in\mathbb{Z}\}$.
\end{remark}

The statement of the following proposition immediately follows from Cor. \ref{degRij}:
\begin{proposition}
Let $P\in\mathcal{L}^{\mathbb{N}}$ be an arbitrary nonzero sequence, let $i,j\in\mathbb{Z}$, $i\neq 0$. 
Then the orbit  $\langle\mathcal{R}_i^j\rangle\cdot P$ is an infinite linearly independent set. In particular, the set $\langle\mathcal{R}_1^l\rangle\cdot P$  is an infinite linearly independent set  for each fixed $ l\in\mathbb{Z}$.
\end{proposition}

Since $\mathcal{L}^{\mathbb{N}}\subset\mathcal{A}^{\mathbb{N}}$, we can immediately conclude one general result for the group $\langle\mathcal{R}^q,\mathcal{R}^m\rangle$:
\begin{corollary}
Each group element $\mathcal{R}_i^j$ with $i,j\in\mathbb{Z},\ i\neq 0$ is of the infinite order, i.e. there is no nonzero $k\in\mathbb{Z}$ such that $(\mathcal{R}_i^j)^k=\mathcal{I}$.
\end{corollary}

\section{$\Bbb Z$-graded Lie algebra of the $\tau^W_{\mathcal L}$-symmetries of the rYME}

As it has already been discussed in \cite{jahn_voj}, the full-fledged recursion operators $\mathcal R_i^j$, $i,j \in \Bbb Z$, provide us with a very pleasant tool for generating (infinite hierarchies of) full-fledged $\tau^W$-symmetries of the rYME \eqref{yme}, since they enable us to avoid lots of tedious computations related to the lifting of shadows of symmetries to their full forms. So it appears to be very advantageous to generate as many $\tau^W$-symmetries as possible in this simple manner.  

In this section, we will focus on the Lie algebra ${\mathrm{sym}}_{\mathcal{L}}^{{\tau}^W}(\mathcal{E}) \subset\mathcal{L}^{\mathbb{N}}$ of the (generating functions of) $\tau^W$-symmetries whose components are the Laurent polynomial functions of the internal coordinates on $\tilde{\mathcal E}^W$ (we will refer to such $\tau^W$-symmetries as \textit{$\tau_{\mathcal{L}}^W$-symmetries}), and we will attempt to utilize the results of the previous sections to address the problems that very naturally arise here. Firstly,
\begin{itemize}
\item[] \textbf{Problem 1.} \textit{To find the smallest set $\mathrm{Seed}_{\mathcal{L}}^{\tau^W}(\mathcal E)\subset \mathrm{sym}_{\mathcal{L}}^{{\tau}^W}(\mathcal{E})$ of $\tau_{\mathcal{L}}^W$-symmetries such that the vector space $\mathrm{sym}_{\mathcal{L}}^{{\tau}^W}(\mathcal{E})$ can be described as the linear span of the elements of the set $\langle\mathcal{R}^m,\mathcal{R}^q\rangle\cdot \mathrm{Seed}_{\mathcal{L}}^{\tau^W}(\mathcal E)$. 
We will refer to the set $\mathrm{Seed}_{\mathcal{L}}^{\tau^W}(\mathcal E)$ as {\bf the set of the seed generators} of the vector space $\mathrm{sym}_{\mathcal{L}}^{{\tau}^W}(\mathcal{E})$}.
\end{itemize}

If we have successfully solved Problem 1, we can further ask what the hierarchies of $\tau^W$-symmetries related to the set of the seed generators look like, that is
\begin{itemize}
\item[] \textbf{Problem 2.} \textit{To study and describe the orbits of the seed generators $P\in\mathrm{Seed}_{\mathcal{L}}^{\tau^W}(\mathcal E)$ under the action of the group $\langle\mathcal{R}^q,\mathcal{R}^m\rangle$ on the set $\mathrm{sym}_{\mathcal{L}}^{\tau^W}(\mathcal{E})$.}
\end{itemize}

The aim of this section is to find at least partial solutions to both of these problems. To this end, we take into account the fact that, using the weight setting \eqref{weightchoice}, the Lie algebra ${\mathrm{sym}}_{\mathcal{L}}^{{\tau}^W}(\mathcal{E})$ becomes $\mathbb{Z}$-graded, i.e. ${\mathrm{sym}}_{\mathcal{L}}^{{\tau}^W}(\mathcal{E})=\bigoplus_{k\in\mathbb{Z}}{\mathrm{sym}}_{\mathcal{L},k}^{{\tau}^W}$, the $\mathbb{Z}$-grading being inherited from the  $\mathbb{Z}$-grading on $\mathcal{L}^{\mathbb{N}}$ (see Sect. \ref{propRO}), and that all the mappings $\mathcal{R}_i^j$ 
leave the set $\mathrm{sym}_{\mathcal{L}}^{\tau^W}(\mathcal{E})$ invariant. Thus, we can deduce that all the results of the previous section can be naturally transferred  to the space ${\mathrm{sym}}_{\mathcal{L}}^{{\tau}^W}(\mathcal{E})$. In particular, the following corollary arises as the consequence of Prop. \ref{grad01} (note that the symbol $\mathcal{R}_i^j$ denotes both the endomorphism on $\mathcal{L}^{\mathbb{N}}$ and its restriction to $\mathrm{sym}_{\mathcal{L}}^{\tau^W}(\mathcal{E}) \subset \mathcal L^{\Bbb N}$).

\begin{corollary}
Upon the weight setting\eqref{weightchoice}, the real vector space {isomorphism} $\mathcal{R}_i^j:{\mathrm{sym}}_{\mathcal{L}}^{{\tau}^W}(\mathcal{E}) \to$ $ {\mathrm{sym}}_{\mathcal{L}}^{{\tau}^W}(\mathcal{E})$ is a graded isomorphism of degree $i$ for each fixed couple $i,j\in\mathbb{Z}$. Consequently, all the vector subspaces ${\mathrm{sym}}^{{\tau}^W}_{\mathcal{L},k}(\mathcal{E})$, where $ k \in \Bbb Z,$ are mutually isomorphic.
\end{corollary}

Therefore, we can conclude that to describe the Lie algebra ${\mathrm{sym}}_{\mathcal L}^{{\tau}^W}(\mathcal{E}) = \bigoplus_{k\in \Bbb Z} {\mathrm{sym}}_{{\mathcal L},k}^{{\tau}^W}(\mathcal{E})$ it is sufficient, in principle, to describe by direct computations only the subspace ${\mathrm{sym}}_{{\mathcal L,0}}^{{\tau}^W}(\mathcal{E})$ since any other subspace ${\mathrm{sym}}_{{\mathcal L,k}}^{{\tau}^W}(\mathcal{E})$, $k \in \Bbb Z$, can be subsequently obtained as the image of the subspace ${\mathrm{sym}}_{{\mathcal L,0}}^{{\tau}^W}(\mathcal{E})$ under the action of the recursion operator $\mathcal R_k^0$.
However, in order to describe the subspace ${\mathrm{sym}}_{{\mathcal L,0}}^{{\tau}^W}(\mathcal{E})$, we would need to solve Eqs. \eqref{nsym-yme1} - \eqref{nsym-yme9} assuming $P=\left[p_0, p_0^q, p_0^m, p_1^r,p_1^q, p_1^m, p_2^r,p_2^q, p_2^m\ldots  \right]\in\mathcal{L}_0^{\mathbb{N}}$, and find a set of generators of  ${\mathrm{sym}}_{{\mathcal L,0}}^{{\tau}^W}(\mathcal{E})$. Although the condition that $P\in\mathcal{L}_0^{\mathbb{N}}$ provides us with a useful constraint for the initial setup of direct computations using the computer algebra system \textit{Jets} \cite{Jets}, we still face up to many hard-to-overcome uncertainties:
\begin{itemize}
\item[1)] As usual, we have no restriction on the jet order of the variables which the components of $P$ may depend on.
\item[2)] The covering $\tau^W$ is infinite-dimensional, but we are able to calculate directly only with a finite number of nonlocal variables.
\item[3)] We have no restriction at our disposal on the degrees of polynomials we work with.
\end{itemize}

\subsection{The Lie subalgebra $V$ of the Lie algebra $\mathrm{sym}_{\mathcal{L}}^{\tau^W}(\mathcal E)$}

All of the above practically means that by direct computations we are always able to find just a vector subspace $V_0$ of the vector space ${\mathrm{sym}}_{{\mathcal L,0}}^{{\tau}^W}(\mathcal{E})$, and, unfortunately, we have no tool at our hand to prove that $V_0=\mathrm{sym}_{{\mathcal L,0}}^{{\tau}^W}(\mathcal{E})$. 

As a result of a large number of calculations, we have obtained the vector space $V_0$ that can be described as the sum of several vector subspaces:
$$V_0:=V_{\Xi,0}+V_{\Upsilon,0}+V_{\Theta,0}+V_{\Omega,0}+V_{\Psi,0},$$
the particular subspaces $V_{\Xi,0}$, $V_{\Upsilon,0}$, etc. being treated in the text below. Based on Cor. \ref{degRij}, we then define the subspace $V=\bigoplus\limits_{k\in\mathbb{Z}}\mathcal{R}_k^0(V_0)\subset\mathrm{sym}_{\mathcal{L}}^{\tau^W}$.

Since one of our tasks (Problem 1) is to find the set of seed generators for $V$, we will try to reduce the set of generators of each of the vector subspaces $V_{\Xi,0}$, $V_{\Upsilon,0}$, etc., thereby ultimately reducing the set of generators of the entire $V_0$ and hence of the entire $V$.

\subsubsection{The vector spaces $V_{\Omega,0}$ and $V_{\Psi,0}$}
\label{v_0m}
In \cite{jahn_voj},  we have presented, among others, the following two examples of  the $\tau_{\mathcal{L}}^W$-symmetries of the rYME \eqref{yme}:
\begin{align*}
\Omega &= \left[tu_t-xu_x+2u,\ldots, tq_{\alpha,t}-xq_{\alpha,x}-\alpha q_\alpha,\right.\\ 
&\hspace{1cm} \left. tm_{\alpha,t}-xm_{\alpha,x}, tr_{\alpha+1,t}-xr_{\alpha+1,x}+(\alpha+3)r_{\alpha+1}, \ldots  \right],
\intertext{resp.}
\Psi&= [u_z, \ldots, q_{\alpha,z}, m_{\alpha,z}-(\alpha+1)m_{\alpha+1}, r_{\alpha+1,z}, \ldots], \qquad \alpha\geq 0.
\end{align*}

It can be easily verified that both of them are homogeneous symmetries of the weight 0 with respect to the weight setting  \eqref{weightchoice}.
The vector space $V_{\Omega,0}$, resp. $V_{\Psi,0}$, is defined to be the smallest subspace of $\mathrm{sym}_{\mathcal{L},0}^{\tau^W}$ that contains $\Omega$, resp. $\Psi$, and is  $\mathcal{R}^m$-invariant. In other words, the space $V_{\Omega,0}$, resp. $V_{\Psi, 0}$, is generated by the elements of the orbit $\langle\mathcal{R}^m\rangle\cdot\Omega$, resp. $\langle\mathcal{R}^m\rangle\cdot\Psi$, i.e.
$$V_{\Omega,0}=\llbracket \langle\mathcal{R}^m\rangle\cdot\Omega \rrbracket\ = \llbracket\mathcal{R}_0^k(\Omega) \ |\ k\in\mathbb{Z} \rrbracket, \quad \ \mathrm{resp.}\quad V_{\Psi, 0}=\llbracket \langle\mathcal{R}^m\rangle\cdot\Psi \rrbracket\ =\llbracket\mathcal{R}_0^k(\Psi)\ |\ k\in\mathbb{Z} \rrbracket.$$
 
The way to obtain the generators of  the spaces $V_{\Omega,0}$ and $V_{\Psi,0}$ have been more or less described in \cite{jahn_voj}.

\subsubsection{The vector spaces $\mathrm{Fam}_{\Xi}$ and $V_{\Xi,0}$}
\label{fam_xi}
The rYME \eqref{yme} also possesses the $\tau^W$-symmetries of the form
$$\Xi(A) = \left[0, A, 0, 0, \frac{1}{1!}\mathfrak X(A), 0, 0, \frac{1}{2!}\mathfrak X^2(A), 0, 0, \frac{1}{3!}\mathfrak X^3(A), \ldots \right],$$
where $A=A(z,q_0)$ is an arbitrary function and
$$\mathfrak X = t\pd{}{z} + \sum \limits_{i=0}^\infty(i+1)q_{i+1}\pd{}{q_i},$$ $\mathfrak X^k = \underbrace{\mathfrak X \circ \ldots \circ \mathfrak X}_{k-times}$, 
cf. Example 5 in \cite{jahn_voj}. 
In accordance with the focus of this article, we restrict our considerations only to the functions $A$ that lie in the set $\mathfrak{a}:=\left\{A= \sum_{\mu,\nu \in \Bbb Z} a_{\mu\nu} z^\mu q_0^\nu \ |\ a_{\mu\nu} \in \Bbb R\right\}$ of all Laurent polynomial functions of the variables $z$ and $q_0$.

Let us define the vector space
$$\mathrm{Fam}_{\Xi}:=\left\{\Xi(A)\ |\ A\in\mathfrak{a}\right\} \subset \mathrm{sym}_{\mathcal{L}}^{\tau^W}(\mathcal{E}),$$ 
and consider the $\mathbb{Z}$-grading on $\mathrm{sym}_{\mathcal{L}}^{\tau^W}(\mathcal{E})$ given by the weight setting \eqref{weightchoice}.  Since each $A\in\mathfrak{a}$ is obviously a homogeneous function of the zero weight, and the operator $\mathfrak X:\mathcal{L}\to\mathcal{L}$ is a $\mathbb{Z}$-graded vector space homomorphism of degree 1, we can easily conclude that all $\tau^W_\mathcal L$-symmetries $\Xi(A)\in \mathrm{Fam}_{\Xi}$ are homogeneous of the weight zero,  i.e. $\mathrm{Fam}_{\Xi}\subset\mathrm{sym}_{\mathcal{L},0}^{\tau^W}$. In addition to that, the $\mathbb{R}$-linearity of $\mathfrak{X}$ enables us to write each $\tau^W_\mathcal L$-symmetry $\Xi(A)\in \mathrm{Fam}_{\Xi}$ as a linear combination of the ``monomial" $\tau_\mathcal{L}^W$-symmetries $\Xi(z^{\mu}q_0^{\nu})$, i.e.
$$\Xi(A) = \sum \limits_{\mu,\nu \in \Bbb Z} a_{\mu\nu} \Xi(z^\mu q_0^\nu),$$
which yields
\begin{equation}\label{gen_Fam_Xi}\mathrm{Fam}_{\Xi}=\llbracket\Xi(z^{\mu}q_0^{\nu})\ |\ \mu,\nu\in\mathbb{Z}\rrbracket.\end{equation}
Note that all the above generators form a linearly independent set.

Let us define $V_{\Xi,0}\subset \mathrm{sym}_{\mathcal{L},0}^{\tau^W}$ to be the vector space generated by all the $\tau_{\mathcal{L}}^W$-symmetries $\Xi(A)\in\mathrm{Fam}_{\Xi}$ and by the elements of  their orbits $\langle \mathcal{R}^m\rangle\cdot\Xi(A)$ under the action of the group $\langle\mathcal{R}^m\rangle$, i.e.
$$V_{\Xi,0}:=\llbracket\mathcal{R}_0^k(\Xi(A))\ |\ A\in\mathfrak{a},\ k\in\mathbb{Z}\rrbracket.$$
From the linearity of $\mathcal{R}_0^k$ for all $k\in\mathbb{Z}$ and from \eqref{gen_Fam_Xi} we immediately have
$$V_{\Xi,0}=\llbracket\mathcal{R}_0^k(\Xi(z^{\mu}q_0^{\nu}))\ |\ \mu,\nu,k\in\mathbb{Z}\rrbracket.$$

In order to solve Problem 1, we want to  reduce the set of the generators of $V_{\Xi,0}$ as far as possible. To this end let us note, that it can be straightforwardly verified that for all $\mu, \nu \in \Bbb Z$ it holds
\begin{align*}
 \Xi(z^\mu q_0^\nu)&=\mathcal R_0^\mu(\Xi(q_0^\nu)),
\end{align*}
hence, 
for any $k\in\mathbb{Z}$ we have
\begin{align*}
\mathcal{R}_0^k( \Xi(z^\mu q_0^\nu))&=\mathcal R_0^{\mu+k}(\Xi(q_0^\nu)),
\end{align*}
which gives us  the final simplification
$$V_{\Xi,0}=\llbracket\mathcal{R}_0^{\mu}(\Xi(q_0^\nu))\ |\ \mu, \nu\in\mathbb{Z}\rrbracket.$$

\subsubsection{The vector spaces $\mathrm{Fam}_{\Upsilon}$ and $V_{\Upsilon,0}$}
\label{fam_up}
In \cite{jahn_voj}, we have also discussed the set of $\tau^W$-symmetries of the rYME \eqref{yme} which are of the form
$$\Upsilon(B) = \left[B,0,0,\frac{1}{1!}\mathfrak V(B),0,0,-\frac{1}{2!}\mathfrak V^2(B),0,0,\frac{1}{3!}\mathfrak V^3(B),0,0,-\frac{1}{4!}\mathfrak V^4(B),0,0, \ldots \right],$$
where $B=B(t,y)$ is an arbitrary function and
$$\mathfrak V = z \pd{}{t} + x\pd{}{y} - 2u\pd{}{x} + 3r_1\pd{}{u} - \sum \limits_{i=1}(i+3)r_{i+1}\pd{}{r_i}.$$ 

As we are interested only in the $\tau_{\mathcal{L}}^{W}$-symmetries, we restrict our attention here to the set $\mathfrak{b}:=\left\{B = \sum_{\mu,\nu \in \Bbb Z} b_{\mu\nu} t^\mu y^\nu \ |\ b_{\mu\nu}\in\mathbb{R}\right\}$ of all Laurent polynomial functions of the variables $t$ and $y$, and consider the vector space
$$\mathrm{Fam}_{\Upsilon}=\left\{\Upsilon(B)\ |\ B\in\mathfrak{b}\right\} \subset\mathrm{sym}_{\mathcal{L}}^{\tau^W}(\mathcal{E}).$$ 

Upon the weight setting \eqref{weightchoice}, the operator $\mathfrak{V}:\mathcal{L}\to\mathcal{L}$ is a homogeneous $\mathbb{Z}$-graded vector space homomorphism of degree $-1$, which means that 
$\Upsilon(B) \in \mathrm{Fam}_{\Upsilon}$ is homogeneous of the weight $\lambda\in\mathbb{Z}$ if and only if $B\in \mathfrak b$ is homogeneous of the same weight (for brevity we denote the Laurent polynomial function $B$ of the weight $\lambda$ as $B_\lambda$). Taking into account that $|t|=1$ and $|y|=2$, we see that $B_\lambda$ must be of the form
$$B_{\lambda} = \sum \limits_{\substack{\mu,\nu \in \Bbb Z \\ \mu+2\nu = \lambda}} b_{\mu\nu} t^\mu y^\nu = \sum \limits_{\nu \in \Bbb Z} b_{\nu} t^{\lambda-2\nu} y^\nu = \sum \limits_{\nu \in \Bbb Z} b_{\nu} t^{\lambda} \left( \frac{y}{t^2} \right)^\nu,$$
and subsequently the equalities 
$$\Upsilon(B_{\lambda})=\Upsilon\left( \sum \limits_{\nu \in \Bbb Z} b_{\nu} t^{\lambda} \left( \frac{y}{t^2} \right)^\nu\right)= \sum \limits_{\nu \in \Bbb Z} b_{\nu} \Upsilon\left(t^{\lambda} \left( \frac{y}{t^2} \right)^\nu\right)$$ 
follow from the linearity of the operator $\mathfrak{V}$. Thus, we can conclude that the vector space $\mathrm{Fam}_{\Upsilon,\lambda} \subset \mathrm{Fam}_{\Upsilon}$ of the homogeneous symmetries $\Upsilon(B_{\lambda}) \in \mathrm{Fam}_{\Upsilon}$ is generated by the set 
$$\left\{\Upsilon\left(t^{\lambda} \left( \frac{y}{t^2} \right)^\nu\right)\ |\ \nu\in\mathbb{Z}\right\}.$$
In particular, for the vector space $\mathrm{Fam}_{\Upsilon, 0}$, we have 
\begin{equation}\label{genB0}\mathrm{Fam}_{\Upsilon, 0}=\Big\llbracket \Upsilon\left(\left( \frac{y}{t^2} \right)^\nu\right)\ |\ \nu\in\mathbb{Z}\Big\rrbracket.\end{equation}

Finally, let us define $V_{\Upsilon,0}\subset\mathrm{sym}_{\mathcal{L},0}^{\tau^W}$ to be the vector space generated by the $\tau^W_{\mathcal{L}}$ - symmetries
$\Upsilon(B_0)\in\mathrm{Fam}_{\Upsilon,0}$ and by all the elements of their orbits under the action of the group $\langle\mathcal{R}^m\rangle$ on the set $\mathrm{sym}^{\tau^W}_{\mathcal{L}}(\mathcal{E})$,
i.e.
$$V_{\Upsilon,0}:=\llbracket \mathcal{R}^k_0(\Upsilon(B_0))\ |\ B_0\in\mathfrak{b},\ k\in\mathbb{Z}\rrbracket.$$

Note that from the $\mathbb{R}$-linearity of $\mathcal{R}_0^k$ we have
$$\mathcal{R}_0^k\left(\Upsilon(B_0)\right)=\mathcal{R}_0^k\left(\sum \limits_{\nu \in \Bbb Z} b_\nu \Upsilon\left(\left( \frac{y}{t^2} \right)^\nu \right)\right)=\sum \limits_{\nu \in \Bbb Z} b_\nu \mathcal{R}_0^k\Upsilon\left(\left( \frac{y}{t^2} \right)^\nu \right).$$
Thus, the set of generators of $V_{\Upsilon,0}$ can be slightly reduced, and we obtain
$$V_{\Upsilon,0}=\Big\llbracket \mathcal{R}^k_0\left(\Upsilon\left(\left( \frac{y}{t^2} \right)^\nu \right)\right)\ |\ k,\nu\in\mathbb{Z}\Big\rrbracket.$$

\subsubsection{The vector spaces $\mathrm{Fam}_{\Theta}$ and $V_{\Theta,0}$}
\label{fam_th}
It  has been shown in \cite{jahn_voj} that  the rYME \eqref{yme} possesses also the symmetries of the form
$$\Theta(C) = \left[0,0,C, 0,0, \frac{1}{1 !}\mathfrak T(C),0,0, \frac{1}{2 !}\mathfrak T^2(C),0,0,\frac{1}{3 !}\mathfrak T^3(C), 0,0,\ldots  \right],$$
where $C=C(z/t,m_0)$ is an arbitrary function of the variables $z/t$ and $m_0$, and
$$\mathfrak T = \pd{}{z}+\sum\limits_{i=0}^\infty (i+1) m_{i+1}\pd{}{m_i}.$$

Similarly to the previous cases, let us define the set $\mathfrak{c}:=\left\{C= \sum_{\mu,\nu \in \Bbb Z} c_{\mu\nu} (z/t)^\mu m_0^\nu\ |\ c_{\mu\nu}\in\mathbb{R}\right\}$
and the vector space
$$\mathrm{Fam}_{\Theta}:=\left\{\Theta(C)\ |\ C\in\mathfrak{c}\right\} \subset\mathrm{sym}_{\mathcal{L}}^{\tau^W}(\mathcal{E}).$$

Upon the weight setting \eqref{weightchoice}, we have $|z/t| = -1$, $|m_\alpha|=0$ for all $\alpha \geq 0$, and the operator $\mathfrak T:\mathcal{L}\to\mathcal{L}$ is a $\mathbb{Z}$-graded vector space homomorphism of the degree $0$. Therefore, it can be easily verified that  $\Theta(C) \in \mathrm{Fam}_\Theta$ is homogeneous of the weight $|\Theta(C)|=\mu$ if and only if $C \in \mathfrak c$ is homogeneous of the weight $|C|=\mu$, i.e.
$$C_{\mu} = (z/t)^{-\mu} \sum \limits_{\nu \in \Bbb Z} c_{\mu\nu} m_0^\nu.$$
For brevity, we denote the Laurent polynomial function $C$ with the weight $\mu$ as $C_{\mu}$ below.

Further, from the linearity of the operator $\mathfrak{T}$, it follows
$$\Theta(C_{\mu})=\Theta\left((z/t)^{-\mu} \sum \limits_{\nu \in \Bbb Z} c_{\mu\nu} m_0^\nu\right)= \sum \limits_{\nu \in \Bbb Z} c_{\mu\nu} \Theta\left((z/t)^{-\mu} m_0^\nu\right)$$
for each homogeneous polynomial $C_\mu\in\mathfrak{c}$. Thus,
we can readily conclude that the vector space $\mathrm{Fam}_{\Theta,\mu}\subset \mathrm{Fam}_\Theta$ of the homogeneous symmetries $\Theta(C_{\mu})  \in \mathrm{Fam}_\Theta$ is generated by the set of symmetries $$\left\{\Theta\left((z/t)^{-\mu} m_0^\nu\right)\ |\ \nu\in\mathbb{Z}\right\}.$$
In particular, for the vector space $\mathrm{Fam}_{\Theta, 0}$ we have 
\begin{equation}
\label{genC0}
\mathrm{Fam}_{\Theta, 0}=\llbracket \Theta\left(m_0^\nu\right)\ |\ \nu\in\mathbb{Z}\rrbracket.
\end{equation}

Finally, let us define $V_{\Theta,0}\subset\mathrm{sym}_{\mathcal{L},0}^{\tau^W}$ to be the vector space generated by the $\tau^W_{\mathcal{L}}$ - symmetries
$\Theta(C_0)\in\mathrm{Fam}_{\Theta,0}$ and by all the elements of their orbits under the action of the group $\langle\mathcal{R}^m\rangle$ on the set $\mathrm{sym}^{\tau^W}(\mathcal{E})$,
i.e.
$$V_{\Theta,0}:=\llbracket \mathcal{R}^k_0(\Theta(C_0))\ |\ C_0\in\mathfrak{c},\ k\in\mathbb{Z}\rrbracket.$$

From the $\mathbb{R}$-linearity of $\mathcal{R}_0^k$ we have
$$\mathcal{R}_0^k\left(\Theta(C_0)\right)=\mathcal{R}^k_0\left(\sum_{\nu\in\mathbb{Z}}c_{\nu}m_0^{\nu}\right)=\sum_{\nu\in\mathbb{Z}}c_{\nu}\mathcal{R}_0^k(\Theta(m_0^{\nu})),$$
thus, the set of generators of $V_{\Theta,0}$ can be slightly reduced, which yields
$$V_{\Theta,0}=\llbracket \mathcal{R}^k_0(\Theta(m_0^\nu))\ |k,\nu\in\mathbb{Z}\rrbracket.$$

\subsubsection{ The set of seed generators and the Lie algebra structure of the space $V$}
Taken together, the results of Sects. \ref{v_0m} - \ref{fam_th} yield
\begin{eqnarray*}V_0&=&V_{\Omega,0}+V_{\Psi,0}+V_{\Xi,0}+V_{\Upsilon,0}+V_{\Theta,0}\\
&=&
\Big\llbracket  \mathcal{R}_0^k(\Omega), \mathcal{R}_0^k(\Psi),\mathcal{R}_0^{k}(\Xi(q_0^\nu)),\mathcal{R}_0^k\left(\Upsilon\left(\left( \frac{y}{t^2} \right)^\nu\right)\right), \mathcal{R}_0^k(\Theta(m_0^{\nu}) \ |\ k\in\mathbb{Z}\Big\rrbracket\\
&=& \Big\llbracket\langle\mathcal{R}^m\rangle\cdot\left\{\Omega, \Psi,\Xi(q_0^{\nu}),\ \Upsilon\left(\left( \frac{y}{t^2} \right)^\nu\right),\ \Theta(m_0^{\nu})\ |\ \nu\in\mathbb{Z} \right\}\Big\rrbracket,
\end{eqnarray*}
and
\begin{equation}
\label{lie_V} 
V=\bigoplus\limits_{k\in\mathbb{Z}}\mathcal{R}_k^0\left(V_0\right)= \Big\llbracket\langle\mathcal{R}^q, \mathcal{R}^m\rangle\cdot\left\{\Omega, \Psi,\Xi(q_0^{\nu}),\ \Upsilon\left(\left( \frac{y}{t^2} \right)^\nu\right),\ \Theta(m_0^{\nu})\ |\ \nu\in\mathbb{Z} \right\}\Big\rrbracket.
\end{equation}
Thus, the desired set of seed generators of $V$ is
 $$\mathrm{Seed}_{\mathcal{L}}^{\tau^W}(V)=\left\{\Omega, \Psi,\Xi(q_0^{\nu}),\ \Upsilon\left(\left( \frac{y}{t^2} \right)^\nu\right),\ \Theta(m_0^{\nu})\ |\ \nu\in\mathbb{Z} \right\},$$
which solves Problem 1 for $V$.

Note that the subspace $V\subset \sym_\mathcal L^{\tau^W}(\mathcal E)$ was constructed only from the zero-weight symmetries. However, we have seen in Sects. \ref{fam_up} and \ref{fam_th} that the vector spaces $\mathrm{Fam}_{\Upsilon}$ and $\mathrm{Fam}_{\Theta}$ contain symmetries of non-zero weights as well. Therefore, it is very natural to ask at this point whether $V$ contains all the symmetries from $\mathrm{Fam}_{\Upsilon}$ and $\mathrm{Fam}_{\Theta}$. 
 Nevertheless, it can be proven that
for all $\lambda, \nu\in\mathbb{Z}$ the equalities
\begin{align*}
\Theta\left(\left(\frac{z}{t}\right)^{-\lambda}m_0^{\nu}\right)=\mathcal{R}_{\lambda}^0\Theta\left(\left(m_0^{\nu}\right)\right),\qquad
\Upsilon\left(t^{\lambda}\left(\frac{y}{t^2}\right)^{\nu}\right)=\mathcal{R}_{\lambda}^{\lambda}\left(\Upsilon\left(\frac{y}{t^2}\right)^{\nu}\right)
\end{align*}
are true, which
 means that all the generators of the vector spaces $\mathrm{Fam}_{\Upsilon}$ and $\mathrm{Fam}_{\Theta}$ lie in V, thus we have $\mathrm{Fam}_{\Upsilon}\subset V$ and $\mathrm{Fam}_{\Theta}\subset V$ as well.

Moreover, it turns out that the vector space $V\subset\mathrm{sym}_{\mathcal{L}}^{\tau^W}(\mathcal{E})$ is closed under the Jacobi bracket, see Tab. \ref{tab_jb}.

\begin{remark}
\label{notation}
In Tab. \ref{tab_jb}, the following notation is used: Given a  $\tau^W_{\mathcal L}$-symmetry $P$ and the hierarchy of symmetries $\langle\mathcal{R}^q,\ \mathcal{R}^m\rangle\cdot P=\left\{\mathcal{R}_i^j(P)\ |\ i,j\in\mathbb{Z}\right\}$ related to $P$, the elements $\mathcal{R}_i^j(P)$ are denoted as $P_i^j$. 
For instance, the elements of the hierarchy related to the seed symmetry $\Xi(A)$ with $A\in\mathfrak{a}$ fixed are denoted as $\Xi_i^j(A) \equiv \mathcal{R}_i^j\left(\Xi(A)\right)$. 
\end{remark}

Regarding the properties of the vector space $V$, we arrive at the following proposition.
\begin{proposition}
The vector space $V=\bigoplus_{k\in\mathbb{Z}}\mathcal{R}_k^0(V_0)$ forms an infinite-dimensional Lie subalgebra of the Lie algebra $\mathrm{sym}_{\mathcal{L}}^{\tau^W}(\mathcal{E})$. The set 
$$\left\{\mathcal{R}_j^k(\Omega), \mathcal{R}_j^k(\Psi),\mathcal{R}_j^{k}(\Xi(q_0^\nu)),\mathcal{R}_j^k\left(\Upsilon\left(\left( \frac{y}{t^2} \right)^\nu\right)\right), \mathcal{R}_j^k(\Theta(m_0^{\nu}) \ |\ j,\nu\in\mathbb{Z}\right\}$$
(which is a subset of the generators set of $V$) is linearly independent for each fixed $k\in\mathbb{Z}$.
\end{proposition}
\begin{proof}
To prove that $V$ is a Lie algebra, it is enough to verify that $V$ is closed under the Jacobi bracket. This can be seen from Tab. \ref{tab_jb}, which was created based on tedious straightforward calculations.

To prove the second part of the assertion, note that the set 
$$\left\{\Omega, \Psi, \Xi(z^{\mu}q_0^{\nu}), \Upsilon\left(\left( \frac{y}{t^2} \right)^\nu\right), \Theta(m_0^{\nu})\ |\ \mu,\nu\in\mathbb{Z}\right\}\subset\mathrm{sym}_{\mathcal{L},0}^{\tau^W}$$ 
is linearly independent. Since for any $j,k\in\mathbb{Z}$ the operator $\mathcal{R}_j^k$ is a vector space isomorphism,  we can conclude that the set $$\left\{\mathcal{R}_j^k(\Omega), \mathcal{R}_j^k(\Psi),\mathcal{R}_j^{k}(\Xi(q_0^\nu)),\mathcal{R}_j^k\left(\Upsilon\left(\left( \frac{y}{t^2} \right)^\nu\right)\right), \mathcal{R}_j^k(\Theta(m_0^{\nu}) \ |\ \nu\in\mathbb{Z}\right\}\subset\mathrm{sym}_{\mathcal{L},j}^{\tau^W}$$
is linearly independent as well. 
Finally, the vector space $\mathrm{sym}_{\mathcal{L}}^{\tau^W}(\mathcal E)$ is the direct sum of the  homogeneous spaces  $\mathrm{sym}_{\mathcal{L},j}^{\tau^W}(\mathcal E)$, which immediately proves the linear independence of the set
$$\left\{\mathcal{R}_j^k(\Omega), \mathcal{R}_j^k(\Psi),\mathcal{R}_j^{k}(\Xi(q_0^\nu)),\mathcal{R}_j^k\left(\Upsilon\left(\left( \frac{y}{t^2} \right)^\nu\right)\right), \mathcal{R}_j^k(\Theta(m_0^{\nu}) \ |\ j, \nu\in\mathbb{Z}\right\}$$ for any fixed $k\in\mathbb{Z}$.
\end{proof}

All of the above, together with the fact that we have found no other $\tau^W_\mathcal L$-symmetries except those contained in $V$  (despite the large number of calculations performed), leads us to formulate the following conjecture.

\begin{conj}
The Lie algebra $V$ defined  by the formula \eqref{lie_V} contains all the $\tau^W_{\mathcal{L}}$-symmetries of the rYME \eqref{yme}, i.e. $V=\mathrm{sym}_{\mathcal{L}}^{\tau^W}(\mathcal{E})$.
\end{conj}

\begin{table}[t]
\resizebox{\textwidth}{!}{
\begin{tblr}{
hline{2}={1pt},
vline{2}={1pt},
colspec={cccccc},
rowsep={5pt}
}
$\{ \downarrow, \rightarrow \}$ &  $\Psi_j^l$ & $\Omega_j^l$ &  $\Xi_j^l(\bar A)$ & $\Upsilon_j^l(\bar B)$ &   $\Theta_j^l(\bar C)$ \\
 $\Psi_i^k$ &  $(k-l)\Psi_{i+j}^{k+l-1}$ &  $i\Psi_{i+j}^{k+l}-l\Omega_{i+j}^{k+l-1}$ & $-\Xi_{i+j}^k((z^l\bar A)_z)$ &  $-\Upsilon_{i+j}^{k+l-1}(l\bar B+t\bar{B}_t)$ & $-l \Theta_{i+j}^{k+l-1}(\bar C)$ \\
$\Omega_i^k$ &  & $(i-j)\Omega_{i+j}^{k+l}$ & $-j\Xi_{i+j}^{k+l}(\bar A)$ & $-\Upsilon_{i+j}^{k+l}((j+2)\bar B+t\bar{B}_t)$ & $\Theta_{i+j}^{k+l}(z \bar C_z-j\bar C)$\\
$\Xi_i^k(A)$ & & & $\Xi_{i+j}^{k+l}(A\bar A_{q_0} - \bar A A_{q_0})$ & $0$ & $0$ \\
$\Upsilon_i^k(B)$ & & & & $\Upsilon_{i+j}^{k+l-2}(t^2(\bar BB_y-B\bar{B}_y))$ & $0$\\
$\Theta_i^k(C)$ & & & & & $\Theta_{i+j}^{k+l}(C \bar C_{m_0}- \bar C C_{m_0})$
\end{tblr}
}
\caption{Jacobi brackets of the $\tau^W$-symmetries of the rYME \eqref{yme}}
\label{tab_jb}
\end{table}

\subsection{The hierarchies related to the seed generators} 
It follows from the results of the previous sections that all (known to us) $\tau^W_{\mathcal{L}}$-symmetries of the rYME \eqref{yme} can be expressed as linear combinations of the elements of the orbits of the $\tau^W_\mathcal L$-symmetries  from the set $\mathrm{Seed}_{\mathcal{L}}^{\tau^W}(V)$
under the group action of $\langle\mathcal{R}^q,\mathcal{R}^m\rangle$. In this section, we examine in detail the properties of elements from selected segments of the set $\langle\mathcal{R}^q,\mathcal{R}^m\rangle\cdot\mathrm{Seed}_{\mathcal{L}}^{\tau^W}(V)$.  It generally holds that each member of the orbit under study that is an invisible $\tau^W_\mathcal L$-symmetry can be described by a general concise formula. On the other hand, if any members of the orbit in question have nontrivial shadows, then, even though these $\tau^W_\mathcal L$-symmetries can still  be computed recursively by direct application of the basic recursion operators (using the formulas \eqref{ROq} - \eqref{InvROm}), we are unfortunately unable to describe them by a general closed formula.

\begin{remark}
Within this section, we mainly use the notation introduced in Rem. \ref{notation}. Since all the hierarchies under study are generated from the homogeneous seed generators of the zero weight (upon the weights setting \eqref{weightchoice}), let us note that the subindex $i \in \Bbb Z$ always coincides (as the consequence of Prop. \ref{grad01}) with the weight of the $\tau^W_\mathcal L$-symmetry $P_i^k$.
\end{remark}

\subsubsection{The hierarchies related to the seed $\tau^W_\mathcal L$-symmetries $\Omega$ and $\Psi$}
\begin{figure}[t]
\begin{center}
\begin{tikzcd}[row sep=0.5cm, column sep=0.4cm]
&
\vdots
&
\vdots
\drar[phantom, "\tau^r", xshift=0.2cm]
&
\vdots
&
\vdots
&
\vdots
&
\vdots
&
\vdots
\drar[phantom, "\tau^q \oplus \tau^r", xshift=0.1cm]
&
\vdots
&
\vdots
\\
\ldots
&
\Omega_{-4}^4
&
\Omega_{-3}^4
&
\Omega_{-2}^4
&
\Omega_{-1}^4
&
\mathbf{\Omega_0^4}
&
\Omega_1^4
&
\Omega_2^4
&
\Omega_3^4
&
\Omega_4^4
&
\ldots
\\
\ldots
&
\Omega_{-4}^3
&
\Omega_{-3}^3
&
\Omega_{-2}^3
&
\Omega_{-1}^3
&
\mathbf{\Omega_0^3}
&
\Omega_1^3
&
\Omega_2^3
&
\Omega_3^3
&
\Omega_4^3
\drar[phantom, "\tau^q"]
&
\ldots
\\
\ldots
&
\Omega_{-4}^2
&
\Omega_{-3}^2
&
\Omega_{-2}^2
&
\Omega_{-1}^2
&
\mathbf{\Omega_0^2}
&
\Omega_1^2
&
\Omega_2^2
&
\Omega_3^2
&
\Omega_4^2
&
\ldots
\\
\ldots
&
\Omega_{-4}^1
&
\Omega_{-3}^1
&
\Omega_{-2}^1
&
\Omega_{-1}^1
&
\mathbf{\Omega_0^1}
\drar[phantom, "local", xshift=0.1cm, yshift=-0.1cm]
&
\Omega_1^1
&
\Omega_2^1
&
\Omega_{3}^1
&
\Omega_{4}^1
&
\ldots
\\
\ldots
\ar[rrrrrrrrrr, bend left=0, no head, yshift=-0.5cm] 
&
\mathbf{\Omega_{-4}^0}
&
\mathbf{\Omega_{-3}^0}
&
\mathbf{\Omega_{-2}^0}
&
\mathbf{\Omega_{-1}^0}
&
\mathbf{\Omega_0^0}
&
\mathbf{\Omega_1^0}
&
\mathbf{\Omega_2^0}
&
\mathbf{\Omega_3^0}
&
\mathbf{\Omega_{4}^0}
&
\ldots
\\
\ldots
\drar[phantom, "\tau^m \oplus \tau^r", xshift=-0.2cm, yshift=-0.1cm]
& 
\Omega_{-4}^{-1}
&
\Omega_{-3}^{-1}
&
\Omega_{-2}^{-1}
& 
\Omega_{-1}^{-1}
& 
\mathbf{\Omega_0^{-1}}
&
\Omega_1^{-1}
&
\Omega_2^{-1}
&
\Omega_3^{-1}
&
\Omega_4^{-1}
&
\ldots
\\
\ldots
& 
\Omega_{-4}^{-2}
&
\Omega_{-3}^{-2}
&
\Omega_{-2}^{-2}
& 
\Omega_{-1}^{-2}
\drar[phantom, "\tau^m"]
& 
\mathbf{\Omega_0^{-2}}
&
\Omega_1^{-2}
&
\Omega_2^{-2}
&
\Omega_3^{-2}
\drar[phantom, "\tau^q \oplus \tau^m"]
&
\Omega_4^{-2}
&
\ldots
\\
&
\vdots
\ar[rurururururururu, bend left=0, no head, xshift=1.0cm, yshift=0.4cm] 
&
\vdots
&
\vdots
&
\vdots
&
\vdots
&
\vdots
\ar[uuuuuuuu, bend left=0, no head, xshift=0.5cm] 
&
\vdots
&
\vdots
&
\vdots
\end{tikzcd}
\end{center}
  \caption{Localization of the $\tau^W_{\mathcal L}$-symmetries from the hierarchy $\left\{ \Omega_i^k\right\}_{i,k \in \Bbb Z}$  by virtue of their shadows}
  \label{fig_omega}
\end{figure}
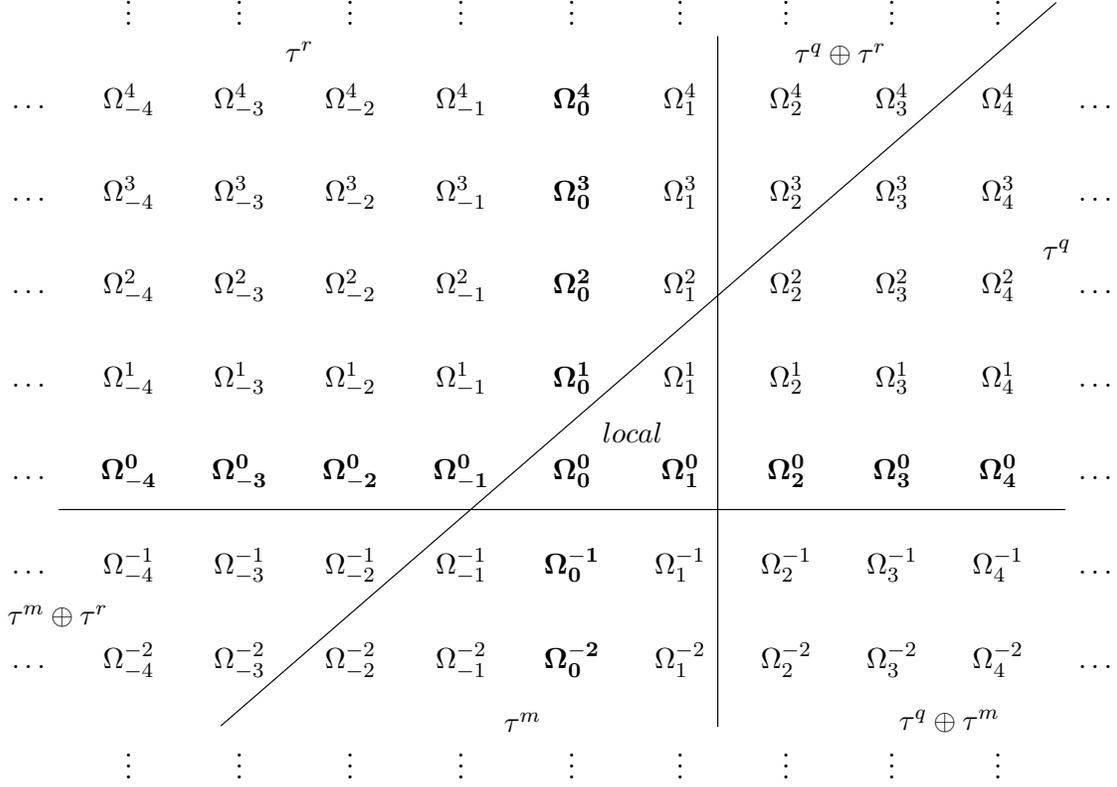

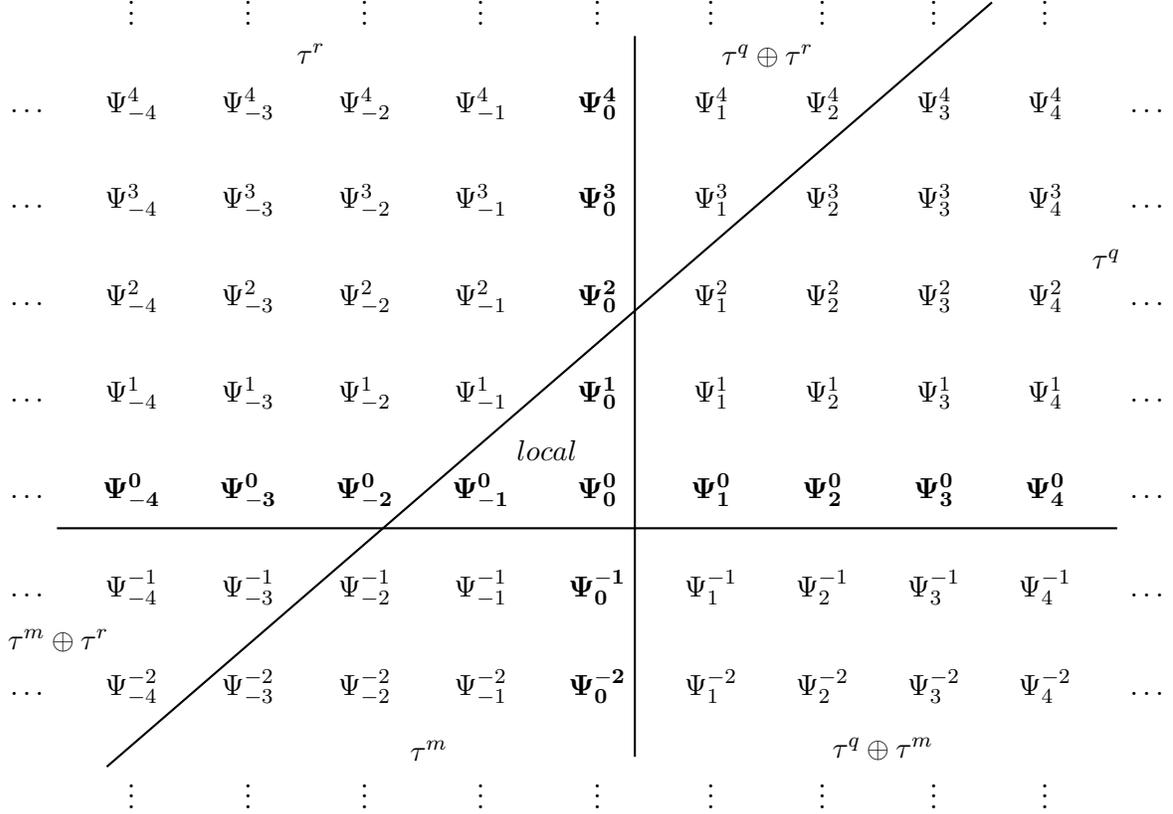
\begin{figure}[t]
\begin{center}
\begin{tikzcd}[row sep=0.55cm, column sep=0.45cm]
&
\vdots
&
\vdots
\drar[phantom, "\tau^r", xshift=0.2cm]
&
\vdots
&
\vdots
&
\vdots
&
\vdots
\drar[phantom, "\tau^q \oplus \tau^r", xshift=0.1cm]
&
\vdots
&
\vdots
&
\vdots
\\
\ldots
&
\Psi_{-4}^4
&
\Psi_{-3}^4
&
\Psi_{-2}^4
&
\Psi_{-1}^4
&
\mathbf{\Psi_0^4}
&
\Psi_1^4
&
\Psi_2^4
&
\Psi_3^4
&
\Psi_4^4
&
\ldots
\\
\ldots
&
\Psi_{-4}^3
&
\Psi_{-3}^3 
&
\Psi_{-2}^3
&
\Psi_{-1}^3
&
\mathbf{\Psi_0^3}
&
\Psi_1^3
&
\Psi_2^3
&
\Psi_3^3
&
\Psi_4^3
\drar[phantom, "\tau^q"]
&
\ldots
\\
\ldots
&
\Psi_{-4}^2
&
\Psi_{-3}^2
&
\Psi_{-2}^2
&
\Psi_{-1}^2
&
\mathbf{\Psi_0^2}
&
\Psi_1^2
&
\Psi_2^2
&
\Psi_3^2
&
\Psi_4^2
&
\ldots
\\
\ldots
&
\Psi_{-4}^1
&
\Psi_{-3}^1
&
\Psi_{-2}^1
&
\Psi_{-1}^1
\drar[phantom, "local", xshift=0.1cm, yshift=-0.1cm]
&
\mathbf{\Psi_0^1}
&
\Psi_1^1
&
\Psi_2^1
&
\Psi_{3}^1
&
\Psi_{4}^1
&
\ldots
\\
\ldots
\ar[rrrrrrrrrr, bend left=0, no head, yshift=-0.5cm, thick] 
&
\mathbf{\Psi_{-4}^0}
&
\mathbf{\Psi_{-3}^0}
&
\mathbf{\Psi_{-2}^0}
&
\mathbf{\Psi_{-1}^0}
&
\mathbf{\Psi_0^0}
&
\mathbf{\Psi_1^0}
&
\mathbf{\Psi_2^0}
&
\mathbf{\Psi_3^0}
&
\mathbf{\Psi_{4}^0}
&
\ldots
\\
\ldots
\drar[phantom, "\tau^m \oplus \tau^r", xshift=-0.2cm, yshift=-0.1cm]
& 
\Psi_{-4}^{-1}
&
\Psi_{-3}^{-1}
&
\Psi_{-2}^{-1}
& 
\Psi_{-1}^{-1}
& 
\mathbf{\Psi_0^{-1}}
&
\Psi_1^{-1}
&
\Psi_2^{-1}
&
\Psi_3^{-1}
&
\Psi_4^{-1}
&
\ldots
\\
\ldots
& 
\Psi_{-4}^{-2}
&
\Psi_{-3}^{-2}
&
\Psi_{-2}^{-2}
\drar[phantom, "\tau^m"]
& 
\Psi_{-1}^{-2}
& 
\mathbf{\Psi_0^{-2}}
&
\Psi_1^{-2}
&
\Psi_2^{-2}
\drar[phantom, "\tau^q \oplus \tau^m"]
&
\Psi_3^{-2}
&
\Psi_4^{-2}
&
\ldots
\\
\ar[rurururururururu, bend left=0, no head, xshift=1.0cm, yshift=0.4cm, thick] 
&
\vdots
&
\vdots
&
\vdots
&
\vdots
&
\vdots
\ar[uuuuuuuu, bend left=0, no head, xshift=0.5cm, thick] 
&
\vdots
&
\vdots
&
\vdots
&
\vdots
\end{tikzcd}
\end{center}
  \caption{Localization of the $\tau^W_{\mathcal L}$-symmetries from the hierarchy $\left\{ \Psi_i^k\right\}_{i,k \in \Bbb Z}$  by virtue of their shadows }
  \label{fig_psi}
\end{figure}

In \cite{jahn_voj}, we have presented, \textit{inter alia}, the shadows of the selected $\tau^W_{\mathcal L}$-symmetries from the hierarchies $\left\{ \Omega_i^k\right\}_{i,k \in \Bbb Z}$ and $\left\{ \Psi_i^k\right\}_{i,k \in \Bbb Z}$, where $\Omega_i^k = \mathcal R_i^k(\Omega)$ and $\Psi_i^k = \mathcal R_i^k(\Psi)$. In our opinion, there is no reason for any explicit listing of others. Instead, we present the schemes of both hierarchies here, see Figs. \ref{fig_omega} and \ref{fig_psi}. 
The orbits of the seed $\tau^W_{\mathcal L}$-symmetries $\Omega \equiv \Omega_0^0$, resp. $\Psi \equiv \Psi_0^0$, with respect to the action of the subgroup $\langle \mathcal R^q \rangle$ are arranged horizontally, the orbits with respect to the action of the subgroup $\langle \mathcal R^m \rangle$ vertically, both of them are highlighted in bold. The vertical, horizontal and diagonal lines divide each diagram into seven sectors that delineate the members of both hierarchies by virtue of their shadows. For instance, the central triangle sector delimits the members with local shadows. The upper-left sector contains the members whose shadows depend nontrivially only on the nonlocal variables $r_\beta$, $\beta \geq 1$, from the covering $\tau^r$, the lower-right sector contains the members whose shadows depend nontrivially on both the nonlocal variables $q_\alpha$ from the covering $\tau^q$ and the nonlocal variables $m_\alpha$ from the covering $\tau^m$, $\alpha \geq 0$, etc. 

\subsubsection{The hierarchy related to the seed $\tau^W_\mathcal L$-symmetry $\Xi(q_0^{\nu})$}

As mentioned in Sec. \ref{fam_xi}, the equality
$$\Xi_0^\mu(q_0^{\nu})=\mathcal{R}_0^\mu\left(\Xi(q_0^{\nu})\right)=\Xi(z^\mu q_0^\nu)$$
holds for all $\mu,\nu\in\mathbb{Z}$, and thus describes the elements of the hierarchy generated by the $\tau_{\mathcal{L}}^W$-symmetry $\Xi(q_0^{\nu})$
that lie on the `vertical line' passing through the seed $\tau^W_\mathcal L$- symmetry $\Xi(q_0^{\nu})$.

The members $\Xi_{i}^\mu(q_0^\nu)=\mathcal{R}_i^\mu\left(\Xi(q_0^{\nu})\right)$, where $i<0$ and $\mu\in\mathbb{Z}$, lying to the left of this `vertical line', can be described by the following closed formula:
\begin{align*}
\Xi_{i}^\mu(q_0^\nu) &= \left[ \smash{\underbrace{0,\ldots, 0}_{-3i+1}}, z^\mu q_0^\nu, 0, 0, \frac{1}{1!}\mathfrak X(z^\mu q_0^\nu), 0, 0, \frac{1}{2!}\mathfrak X^2(z^\mu q_0^\nu), 0, 0, \frac{1}{3!}\mathfrak X^3(z^\mu q_0^\nu), \ldots \right].
\end{align*}

On the other hand, it turns out that all the $\tau^W_{\mathcal{L}}$-symmetries $\Xi_{i}^\mu(q_0^\nu)=\mathcal{R}_i^\mu\left(\Xi(q_0^{\nu})\right)$ with $i>0$ have nontrivial shadows that depend purely on the nonlocal variables from the covering $\tau^q$. As already mentioned, we are not able to describe them with any concise formula, however they can be computed recursively from $\Xi(q_0^{\nu})$ using the formulas \eqref{ROq} - \eqref{InvROm}.

\begin{figure}[t]
\begin{center}
\begin{tikzcd}[row sep=1.2cm, column sep=1.1cm]
\Xi_{-2,7}^2(z^2 q_0^\nu)
\arrow[bend left=15, "\mathcal R_1^0" description]{r} 
\arrow[bend right=20, "\mathcal R_0^{-1}" description]{d}
&
\Xi_{-1,4}^2(z^2 q_0^\nu)
\arrow[bend left=15, "\mathcal R_1^0" description]{r} 
\arrow[bend left=15, "\mathcal R_{-1}^0" description]{l} 
\arrow[bend right=20, "\mathcal R_0^{-1}" description]{d}
&
\mathbf{\Xi_{0,1}^2(z^2 q_0^\nu)}
\arrow[bend left=15, "\mathcal R_1^0" description]{r} 
\arrow[bend right=20, "\mathcal R_0^{-1}" description]{d}
\arrow[bend left=15, "\mathcal R_{-1}^0" description]{l} 
&
\Xi_{1,0}^2\left( \frac{z^2 q_0^\nu}{q_{0,x}} \right)
\arrow[bend left=15, "\mathcal R_{-1}^0" description]{l} 
\arrow[bend right=20, "\mathcal R_0^{-1}" description]{d}
\arrow[bend left=15, "\mathcal R_1^0" description]{r} 
&
\Xi_{2,0}^2( A_2^2)
\arrow[bend left=15, "\mathcal R_{-1}^0" description]{l} 
\arrow[bend right=20, "\mathcal R_0^{-1}" description]{d}
\\
\Xi_{-2,7}^1(z q_0^\nu)
\arrow[bend left=15, "\mathcal R_1^0" description]{r} 
\arrow[bend right=20, "\mathcal R_0^{-1}" description]{d}
\arrow[bend right=20, "\mathcal R_0^1" description]{u}
&
\Xi_{-1,4}^1(z q_0^\nu)
\arrow[bend left=15, "\mathcal R_1^0" description]{r} 
\arrow[bend left=15, "\mathcal R_{-1}^0" description]{l} 
\arrow[bend right=20, "\mathcal R_0^{-1}" description]{d}
\arrow[bend right=20, "\mathcal R_0^1" description]{u}
&
\mathbf{\Xi_{0,1}^1(zq_0^\nu)}
\arrow[bend left=15, "\mathcal R_1^0" description]{r} 
\arrow[bend right=20, "\mathcal R_0^{-1}" description]{d} 
\arrow[bend right=20, "\mathcal R_0^1" description]{u}
\arrow[bend left=15, "\mathcal R_{-1}^0" description]{l} 
&
\Xi_{1,0}^1\left( \frac{z q_0^\nu}{q_{0,x}} \right)
\arrow[bend left=15, "\mathcal R_{-1}^0" description]{l} 
\arrow[bend right=20, "\mathcal R_0^{-1}" description]{d}
\arrow[bend right=20, "\mathcal R_0^1" description]{u}
\arrow[bend left=15, "\mathcal R_1^0" description]{r} 
&
\Xi_{2,0}^1( A_2^1)
\arrow[bend right=20, "\mathcal R_0^{-1}" description]{d}
\arrow[bend right=20, "\mathcal R_0^1" description]{u}
\arrow[bend left=15, "\mathcal R_{-1}^0" description]{l} 
\\ 
\Xi_{-2,7}^0(q_0^\nu)
\arrow[bend left=15, "\mathcal R_1^0" description]{r} 
\arrow[bend right=20, "\mathcal R_0^{-1}" description]{d}
\arrow[bend right=20, "\mathcal R_0^1" description]{u}
&
\Xi_{-1,4}^0(q_0^\nu)
\arrow[bend left=15, "\mathcal R_1^0" description]{r} 
\arrow[bend left=15, "\mathcal R_{-1}^0" description]{l} 
\arrow[bend right=20, "\mathcal R_0^{-1}" description]{d}
\arrow[bend right=20, "\mathcal R_0^1" description]{u}
&
\mathbf{\Xi_{0,1}^0(q_0^\nu)}
\arrow[bend left=15, "\mathcal R_1^0" description]{r} 
\arrow[bend right=20, "\mathcal R_0^1" description]{u}
\arrow[bend right=20, "\mathcal R_0^{-1}" description]{d}
\arrow[bend left=15, "\mathcal R_{-1}^0" description]{l} 
&
\Xi_{1,0}^0\left( \frac{q_0^\nu}{q_{0,x}} \right)
\arrow[bend left=15, "\mathcal R_{-1}^0" description]{l} 
\arrow[bend right=20, "\mathcal R_0^1" description]{u}
\arrow[bend right=20, "\mathcal R_0^{-1}" description]{d}
\arrow[bend left=15, "\mathcal R_1^0" description]{r} 
&
\Xi_{2,0}^0( A_2^0)
\arrow[bend right=20, "\mathcal R_0^1" description]{u}
\arrow[bend left=15, "\mathcal R_{-1}^0" description]{l} 
\arrow[bend right=20, "\mathcal R_0^{-1}" description]{d}
\\
\Xi_{-2,7}^{-1}\left( \frac{q_0^\nu}{z} \right)
\arrow[bend left=15, "\mathcal R_1^0" description]{r} 
\arrow[bend right=20, "\mathcal R_0^{-1}" description]{d}
\arrow[bend right=20, "\mathcal R_0^1" description]{u}
&
\Xi_{-1,4}^{-1}\left( \frac{q_0^\nu}{z} \right)
\arrow[bend left=15, "\mathcal R_1^0" description]{r} 
\arrow[bend left=15, "\mathcal R_{-1}^0" description]{l} 
\arrow[bend right=20, "\mathcal R_0^{-1}" description]{d}
\arrow[bend right=20, "\mathcal R_0^1" description]{u}
&
\mathbf{\Xi_{0,1}^{-1}\left( \frac{q_0^\nu}{z} \right)}
\arrow[bend right=20, "\mathcal R_0^1" description]{u}
\arrow[bend right=20, "\mathcal R_0^{-1}" description]{d}
\arrow[bend left=15, "\mathcal R_{-1}^0" description]{l} 
\arrow[bend left=15, "\mathcal R_1^0" description]{r} 
&
\Xi_{1,0}^{-1}\left( \frac{q_0^\nu}{zq_{0,x}} \right)
\arrow[bend left=15, "\mathcal R_{-1}^0" description]{l} 
\arrow[bend right=20, "\mathcal R_0^{-1}" description]{d}
\arrow[bend right=20, "\mathcal R_0^1" description]{u}
\arrow[bend left=15, "\mathcal R_1^0" description]{r} 
&
\Xi_{2,0}^{-1}( A_2^{-1})
\arrow[bend left=15, "\mathcal R_{-1}^0" description]{l} 
\arrow[bend right=20, "\mathcal R_0^{-1}" description]{d}
\arrow[bend right=20, "\mathcal R_0^1" description]{u}
\\
\Xi_{-2,7}^{-2}\left( \frac{q_0^\nu}{z^2} \right)
\arrow[bend left=15, "\mathcal R_1^0" description]{r} 
\arrow[bend right=20, "\mathcal R_0^1" description]{u}
&
\Xi_{-1,4}^{-2}\left( \frac{q_0^\nu}{z^2} \right)
\arrow[bend left=15, "\mathcal R_1^0" description]{r} 
\arrow[bend left=15, "\mathcal R_{-1}^0" description]{l} 
\arrow[bend right=20, "\mathcal R_0^1" description]{u}
&
\mathbf{\Xi_{0,1}^{-2}\left( \frac{q_0^\nu}{z^2} \right)}
\arrow[bend left=15, "\mathcal R_{-1}^0" description]{l} 
\arrow[bend left=15, "\mathcal R_1^0" description]{r} 
\arrow[bend right=20, "\mathcal R_0^1" description]{u}
&
\Xi_{1,0}^{-2}\left( \frac{q_0^\nu}{z^2q_{0,x}} \right)
\arrow[bend right=20, "\mathcal R_0^1" description]{u}
\arrow[bend left=15, "\mathcal R_{-1}^0" description]{l} 
\arrow[bend left=15, "\mathcal R_1^0" description]{r} 
&
\Xi_{2,0}^{-2}( A_2^{-2})
\arrow[bend right=20, "\mathcal R_0^1" description]{u}
\arrow[bend left=15, "\mathcal R_{-1}^0" description]{l} 
\end{tikzcd}
\end{center}
  \caption{A part of the orbit of the seed $\tau^W_\mathcal L$-symmetry $\Xi(q_0^\nu)$}
  \label{fig_xi}
\end{figure}
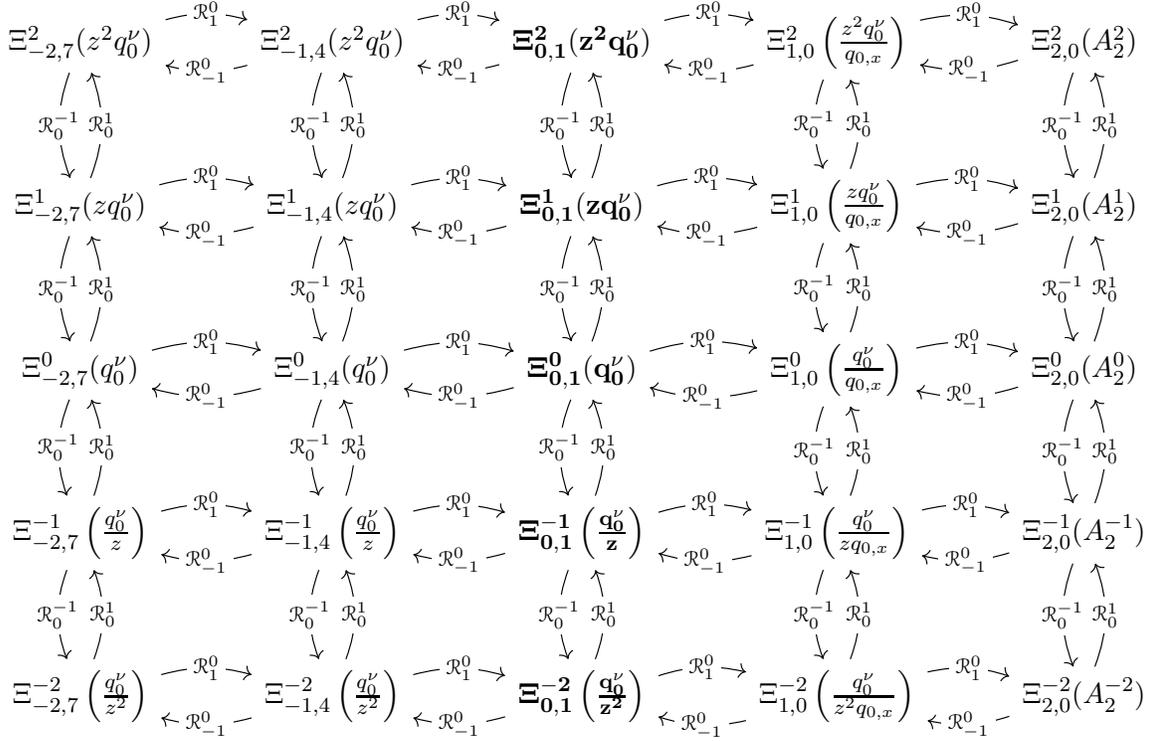

\begin{remark}
\label{orbit_xi}
The structure of the orbit of the seed $\tau^W_\mathcal L$-symmetry  $\Xi(q_0^\nu)$ under the action of the group $\langle R^q, R^m \rangle$ is illustrated in Fig. \ref{fig_xi}. For better visualization, we have slightly modified (in comparison with the preceding exposition) the notation of symmetries inside the figure. The first lower index and the upper index still denote the position of the given symmetry relative to the seed symmetry (as mentioned, the lower index also denotes the weight of the symmetry). The second lower index indicates the number of zeros preceding the first non-trivial component of the symmetry in question. Finally, in parentheses, the first non-trivial component of the symmetry is specified rigorously (unlike the previous notation, where the term in parentheses indicated the function A, which uniquely specified the seed symmetry $\Xi(A)$).
Some of the shadows are listed separately in the following list:
\begin{align*}
A_2^{-2} &= \frac{\nu z q_1 q_0^{\nu-1} - 2tq_0^\nu}{z^3 q_{0,x}} - \frac{q_{1,x} q_0^\nu}{z^2 q_{0,x}^2},  & A_2^{-1} &= \frac{\nu z q_1 q_0^{\nu-1} - tq_0^\nu}{z^2 q_{0,x}} - \frac{q_{1,x} q_0^\nu}{z q_{0,x}^2},\\[2mm] 
A_2^0 &= \frac{\nu q_1 q_0^{\nu-1}}{q_{0,x}} - \frac{q_{1,x} q_0^\nu}{q_{0,x}^2}, & A_2^1 &= \frac{\nu z q_1 q_0^{\nu-1} + tq_0^\nu}{q_{0,x}} - \frac{zq_{1,x} q_0^\nu}{q_{0,x}^2},\\[2mm]
 A_2^2 &= \frac{\nu z^2 q_1 q_0^{\nu-1} + 2tzq_0^\nu}{q_{0,x}} - \frac{z^2 q_{1,x} q_0^\nu}{q_{0,x}^2}.
\end{align*}
\end{remark}
The $\tau^W_\mathcal L$-symmetries $\Xi_{0,1}^j(z^jq_0^{\nu}) = \Xi(z^jq_0^{\nu})$, $j\in\mathbb{Z}$, belong to the vector space $\mathrm{Fam}_{\Xi}$ which has been defined in Sec. \ref{fam_xi}, and they are highlighted in bold.
As we can see, they demarcate the boundary between the area of the invisible $\tau^W_\mathcal L$-symmetries and the area of the $\tau^W_\mathcal L$-symmetries with nontrivial shadows. The crossing between these areas runs in the horizontal direction here, thus it is mediated by the basic recursion operator $\mathcal R_1^0$ and its inversion.

\subsubsection{The hierarchy related to the seed $\tau^W_\mathcal L$-symmetry $\Upsilon\left(\left( \frac{y}{t^2} \right)^\nu \right)$}

Here, we will mention only selected members of this hierarchy, specifically the elements of the orbits $\langle\mathcal{R}^q\rangle\cdot \Upsilon\left(\left( \frac{y}{t^2} \right)^\nu \right)$ (the `horizontal line'),  $\langle\mathcal{R}^m\rangle\cdot \Upsilon\left(\left( \frac{y}{t^2} \right)^\nu \right)$ (the `vertical line') and $\langle\mathcal{R}_1^1\rangle\cdot \Upsilon\left(\left( \frac{y}{t^2} \right)^\nu \right)$ (the `diagonal line' ). 

Let us describe the `vertical line' of $\tau_{\mathcal{L}}^W$-symmetries passing through $\Upsilon\left(\left( \frac{y}{t^2} \right)^\nu \right)$.
It can be verified that the $\tau_{\mathcal{L}}^W$-symmetry $\Upsilon_0^{\mu}((y/t^2)^\nu)$, $\mu \leq -1$, is an invisible symmetry of the form
\begin{align*}
\Upsilon_0^{\mu}&((y/t^2)^\nu) = \mathcal R_0^{\mu} \left(\Upsilon\left(\left(y/t^2 \right)^\nu\right)\right)\\
&=(-1)^{-\mu-1}\left[ \smash{\underbrace{0,\ldots, 0}_{-3\mu}}, t^{\mu}(y/t^2)^\nu, 0, 0, -\frac{1}{1!}\mathfrak V(t^{\mu}(y/t^2)^\nu), 0, 0, \frac{1}{2!}\mathfrak V^2(t^{\mu}(y/t^2)^\nu), 0, 0, \ldots \right].
\end{align*}

In case $\mu\geq 1$, the corresponding $\tau^W_\mathcal L$-symmetry $\Upsilon_0^{\mu} \left(\left( \frac{y}{t^2} \right)^\nu \right)=\mathcal{R}_0^{\mu} (\Upsilon\left(\left( \frac{y}{t^2} \right)^\nu \right))$ is a symmetry with a nontrivial shadow. It can be computed by the direct iterative application of the basic recursion operator $\mathcal{R}^m$ on the symmetry $\Upsilon\left(\left( \frac{y}{t^2} \right)^\nu \right)$, but we are not able to give any concise formula describing it. 

The elements lying on the `horizontal line' passing through the element $\Upsilon\left(\left( \frac{y}{t^2} \right)^\nu \right)$ can be described as follows.
The members $\Upsilon_i^0\left(\left( \frac{y}{t^2} \right)^\nu \right)=\mathcal{R}_{i}^0\left(\Upsilon\left(\left( \frac{y}{t^2} \right)^\nu\right)\right)$, where  $i \leq -1$, are the $\tau_{\mathcal{L}}^W$-symmetries with nontrivial shadows. These shadows, if nonlocal, depend purely on the nonlocal variables from the covering $\tau^r$. 

The members $\Upsilon_i^0\left(\left( \frac{y}{t^2} \right)^\nu \right)$, where $i \geq 1$, are invisible $\tau^W_\mathcal L$-symmetries that can be concisely described by the following formula:
\begin{align*}
\Upsilon_i^0&((y/t^2)^\nu) = \mathcal R_i^0 \left(\Upsilon\left(\left(y/t^2 \right)^\nu\right)\right)\\
&= (-1)^{i-1}\left[ \smash{\underbrace{0,\ldots, 0}_{3i}}, (y/t^2)^\nu, 0, 0, -\frac{1}{1!}\mathfrak V((y/t^2)^\nu), 0, 0, \frac{1}{2!}\mathfrak V^2((y/t^2)^\nu), 0, 0, \ldots \right].
\end{align*}

Finally, it turns out that the `diagonal' terms belong  to the vector space $\mathrm{Fam}_{\Upsilon}$ from Sec. \ref{fam_up}. More precisely, it can be verified that the equality
$$\Upsilon_{\mu}^\mu\left(\left(y/t^2 \right)^\nu\right)=\Upsilon(t^\mu(y/t^2)^{\nu})$$ is true for all $\mu,\nu \in\mathbb{Z}$.

\begin{figure}[t]
\begin{center}
\begin{tikzcd}[row sep=1.2cm, column sep=1.1cm]
\Upsilon_{-2,0}^2(B_{-2}^2)
\arrow[bend left=15, "\mathcal R_1^0" description]{r} 
\arrow[bend right=20, "\mathcal R_0^{-1}" description]{d}
&
\Upsilon_{-1,0}^2(B_{-1}^2)
\arrow[bend left=15, "\mathcal R_1^0" description]{r} 
\arrow[bend left=15, "\mathcal R_{-1}^0" description]{l} 
\arrow[bend right=20, "\mathcal R_0^{-1}" description]{d}
&
\Upsilon_{0,0}^2(B_0^2)
\arrow[bend left=15, "\mathcal R_1^0" description]{r} 
\arrow[bend right=20, "\mathcal R_0^{-1}" description]{d}
\arrow[bend left=15, "\mathcal R_{-1}^0" description]{l} 
&
\Upsilon_{1,0}^2\left( B_1^2 \right)
\arrow[bend left=15, "\mathcal R_{-1}^0" description]{l} 
\arrow[bend right=20, "\mathcal R_0^{-1}" description]{d}
\arrow[bend left=15, "\mathcal R_1^0" description]{r} 
&
\mathbf{\Upsilon_{2,0}^2( t^2s_0^\nu)}
\arrow[bend left=15, "\mathcal R_{-1}^0" description]{l} 
\arrow[bend right=20, "\mathcal R_0^{-1}" description]{d}
\\
\Upsilon_{-2,0}^1(B_{-2}^1)
\arrow[bend left=15, "\mathcal R_1^0" description]{r} 
\arrow[bend right=20, "\mathcal R_0^{-1}" description]{d}
\arrow[bend right=20, "\mathcal R_0^1" description]{u}
&
\Upsilon_{-1,0}^1(B_{-1}^1)
\arrow[bend left=15, "\mathcal R_1^0" description]{r} 
\arrow[bend left=15, "\mathcal R_{-1}^0" description]{l} 
\arrow[bend right=20, "\mathcal R_0^{-1}" description]{d}
\arrow[bend right=20, "\mathcal R_0^1" description]{u}
&
\Upsilon_{0,0}^1(B_0^1)
\arrow[bend left=15, "\mathcal R_1^0" description]{r} 
\arrow[bend right=20, "\mathcal R_0^{-1}" description]{d} 
\arrow[bend right=20, "\mathcal R_0^1" description]{u}
\arrow[bend left=15, "\mathcal R_{-1}^0" description]{l} 
&
\mathbf{\Upsilon_{1,0}^1\left( ts_0^\nu \right)}
\arrow[bend left=15, "\mathcal R_{-1}^0" description]{l} 
\arrow[bend right=20, "\mathcal R_0^{-1}" description]{d}
\arrow[bend right=20, "\mathcal R_0^1" description]{u}
\arrow[bend left=15, "\mathcal R_1^0" description]{r} 
&
\Upsilon_{2,3}^1( ts_0^\nu)
\arrow[bend right=20, "\mathcal R_0^{-1}" description]{d}
\arrow[bend right=20, "\mathcal R_0^1" description]{u}
\arrow[bend left=15, "\mathcal R_{-1}^0" description]{l} 
\\ 
\Upsilon_{-2,0}^0(B_{-2}^0)
\arrow[bend left=15, "\mathcal R_1^0" description]{r} 
\arrow[bend right=20, "\mathcal R_0^{-1}" description]{d}
\arrow[bend right=20, "\mathcal R_0^1" description]{u}
&
\Upsilon_{-1,0}^0(B_{-1}^0)
\arrow[bend left=15, "\mathcal R_1^0" description]{r} 
\arrow[bend left=15, "\mathcal R_{-1}^0" description]{l} 
\arrow[bend right=20, "\mathcal R_0^{-1}" description]{d}
\arrow[bend right=20, "\mathcal R_0^1" description]{u}
&
\mathbf{\Upsilon_{0,0}^0(s_0^\nu)}
\arrow[bend left=15, "\mathcal R_1^0" description]{r} 
\arrow[bend right=20, "\mathcal R_0^1" description]{u}
\arrow[bend right=20, "\mathcal R_0^{-1}" description]{d}
\arrow[bend left=15, "\mathcal R_{-1}^0" description]{l} 
&
\Upsilon_{1,3}^0\left( s_0^\nu \right)
\arrow[bend left=15, "\mathcal R_{-1}^0" description]{l} 
\arrow[bend right=20, "\mathcal R_0^1" description]{u}
\arrow[bend right=20, "\mathcal R_0^{-1}" description]{d}
\arrow[bend left=15, "\mathcal R_1^0" description]{r} 
&
\Upsilon_{2,6}^0( -s_0^\nu)
\arrow[bend right=20, "\mathcal R_0^1" description]{u}
\arrow[bend left=15, "\mathcal R_{-1}^0" description]{l} 
\arrow[bend right=20, "\mathcal R_0^{-1}" description]{d}
\\
\Upsilon_{-2,0}^{-1}\left( B_{-2}^{-1} \right)
\arrow[bend left=15, "\mathcal R_1^0" description]{r} 
\arrow[bend right=20, "\mathcal R_0^{-1}" description]{d}
\arrow[bend right=20, "\mathcal R_0^1" description]{u}
&
\mathbf{\Upsilon_{-1,0}^{-1}\left( \frac{s_0^\nu}{t} \right)}
\arrow[bend left=15, "\mathcal R_1^0" description]{r} 
\arrow[bend left=15, "\mathcal R_{-1}^0" description]{l} 
\arrow[bend right=20, "\mathcal R_0^{-1}" description]{d}
\arrow[bend right=20, "\mathcal R_0^1" description]{u}
&
\Upsilon_{0,3}^{-1}\left( \frac{s_0^\nu}{t} \right)
\arrow[bend right=20, "\mathcal R_0^1" description]{u}
\arrow[bend right=20, "\mathcal R_0^{-1}" description]{d}
\arrow[bend left=15, "\mathcal R_{-1}^0" description]{l} 
\arrow[bend left=15, "\mathcal R_1^0" description]{r} 
&
\Upsilon_{1,6}^{-1}\left( -\frac{s_0^\nu}{t} \right)
\arrow[bend left=15, "\mathcal R_{-1}^0" description]{l} 
\arrow[bend right=20, "\mathcal R_0^{-1}" description]{d}
\arrow[bend right=20, "\mathcal R_0^1" description]{u}
\arrow[bend left=15, "\mathcal R_1^0" description]{r} 
&
\Upsilon_{2,9}^{-1}\left(\frac{s_0^\nu}{t}\right)
\arrow[bend left=15, "\mathcal R_{-1}^0" description]{l} 
\arrow[bend right=20, "\mathcal R_0^{-1}" description]{d}
\arrow[bend right=20, "\mathcal R_0^1" description]{u}
\\
\mathbf{\Upsilon_{-2,0}^{-2}\left( \frac{s_0^\nu}{t^2} \right)}
\arrow[bend left=15, "\mathcal R_1^0" description]{r} 
\arrow[bend right=20, "\mathcal R_0^1" description]{u}
&
\Upsilon_{-1,3}^{-2}\left( \frac{s_0^\nu}{t^2} \right)
\arrow[bend left=15, "\mathcal R_1^0" description]{r} 
\arrow[bend left=15, "\mathcal R_{-1}^0" description]{l} 
\arrow[bend right=20, "\mathcal R_0^1" description]{u}
&
\Upsilon_{0,6}^{-2}\left( -\frac{s_0^\nu}{t^2} \right)
\arrow[bend left=15, "\mathcal R_{-1}^0" description]{l} 
\arrow[bend left=15, "\mathcal R_1^0" description]{r} 
\arrow[bend right=20, "\mathcal R_0^1" description]{u}
&
\Upsilon_{1,9}^{-2}\left( \frac{s_0^\nu}{t^2} \right)
\arrow[bend right=20, "\mathcal R_0^1" description]{u}
\arrow[bend left=15, "\mathcal R_{-1}^0" description]{l} 
\arrow[bend left=15, "\mathcal R_1^0" description]{r} 
&
\Upsilon_{2,12}^{-2}\left( -\frac{s_0^\nu}{t^2}\right)
\arrow[bend right=20, "\mathcal R_0^1" description]{u}
\arrow[bend left=15, "\mathcal R_{-1}^0" description]{l} 
\end{tikzcd}
\end{center}
  \caption{A part of the orbit of the seed $\tau^W_\mathcal L$-symmetry $\Upsilon(s_0^\nu)$}
  \label{fig_upsilon}
\end{figure}
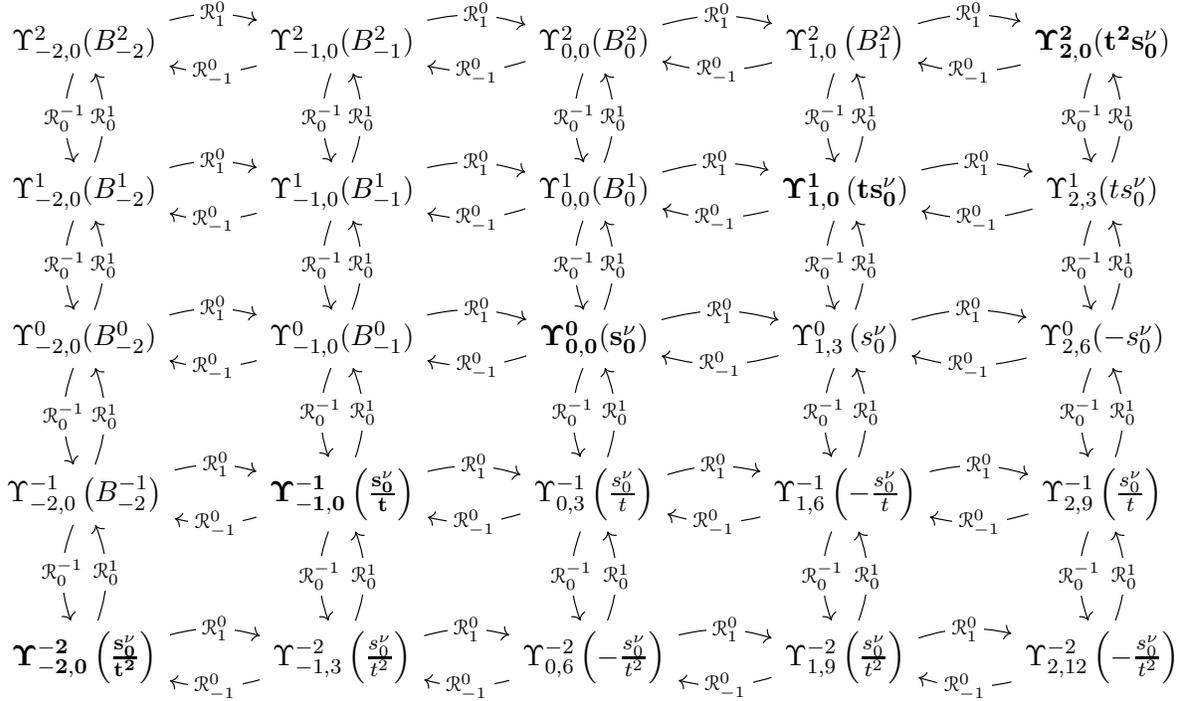

\begin{remark}
\label{orbit_upsilon}
A part of the orbit of the seed $\tau^W_\mathcal L$-symmetry $\Upsilon((y/t^2)^\nu)$ is presented in Fig. \ref{fig_upsilon}. Besides the modification of the notation of the symmetries stated in Rem. \ref{orbit_xi}, we also put for shortness $s_0:=y/t^2$. We present here the explicit forms only of two shadows from the figure in question, namely,
\begin{align*}
B_{-1}^0 &= \left( u_x-\frac{2\nu z}{t} + \frac{\nu x}{y} \right) s_0^\nu,  & B_0^1 &= \left( tu_x- 2\nu z  + \frac{\nu t x}{y} + z \right) s_0^\nu.
\end{align*}
The other ones are omitted here because they are given by too long formulas.
\end{remark}

As mentioned above, the members $\Upsilon_{\mu,0}^\mu(t^\mu s_0^{\nu})=\Upsilon(t^\mu s_0^{\nu})$, where $\mu\in\mathbb{Z}$, belong to the vector space $\mathrm{Fam}_{\Upsilon}$ defined in Sec. \ref{fam_up}. These $\tau^W_\mathcal L$-symmetries are highlighted in bold, and they demarcate the boundary between the area of the invisible $\tau^W_\mathcal L$-symmetries and the area of the $\tau^W_\mathcal L$-symmetries with nontrivial shadows which runs diagonally in this case. This means that the crossing between these areas is realized here by both the basic recursion operators $\mathcal R_1^0$ and $\mathcal R_0^1$ and their inversions.
\begin{remark}
Since the recursion operator $\mathcal{R}_1^1$ maps the set of generators of $\mathrm{Fam}_{\Upsilon,k}$ to the set of generators of  $\mathrm{Fam}_{\Upsilon,k+1}$, we can conclude that  the vector space $\mathrm{Fam}_{\Upsilon}$ is $\mathcal{R}_1^1$-invariant.
\end{remark}

\subsubsection{The hierarchy related to the seed $\tau^W_\mathcal L$-symmetry  $\Theta(m_0^{\nu})$}
Similarly to the previous case, we present here only selected $\tau_{\mathcal{L}}^W$-symmetries from the hierarchy in question, specifically the members lying on the `horizontal line' and on the `vertical line'  passing through the seed  $\tau^W_\mathcal L$-symmetry  $\Theta(m_0^{\nu})$.

 Let us first describe the members of the orbit $\langle\mathcal{R}^m\rangle\cdot\Theta(m_0^{\nu})$, i.e. the elements that lie on the `vertical line'.
It turns out that the members $\Theta_0^i(m_0^\nu)=\mathcal{R}_0^i\left(\Theta_0^i(m_0^\nu)\right)$, where $i>0$, are the invisible $\tau_{\mathcal{L}}^W$-symmetries described by the formula
\begin{align*}
\Theta_0^i(m_0^\nu) &= (-1)^i \left[ \smash{\underbrace{0,\ldots, 0}_{3i+2}}, m_0^\nu, 0, 0, \frac{1}{1!}\mathfrak T(m_0^\nu), 0, 0, \frac{1}{2!}\mathfrak T^2(m_0^\nu), 0, 0, \frac{1}{3!}\mathfrak T^3(m_0^\nu),0,0, \ldots \right].
\end{align*}

The elements that lie on the `vertical line' in the opposite direction, i.e. the $\tau^W_\mathcal L$-symmetries $\Theta_0^{i}(m_0^{\nu})=\mathcal{R}_0^{i}\left(\Theta(m_0^{\nu})\right)$, where $i<0$, 
have nontrivial shadows which depend purely on the nonlocal variables from the covering $\tau^m$. 

\begin{figure}[t]
\begin{center}
\begin{tikzcd}[row sep=1.2cm, column sep=1.2cm]
\Theta_{-2,8}^2\left(\frac{z^2 m_0^\nu}{t^2} \right)
\arrow[bend left=15, "\mathcal R_1^0" description]{r} 
\arrow[bend right=20, "\mathcal R_0^{-1}" description]{d}
&
\Theta_{-1,8}^2\left(-\frac{z m_0^\nu}{t} \right)
\arrow[bend left=15, "\mathcal R_1^0" description]{r} 
\arrow[bend left=15, "\mathcal R_{-1}^0" description]{l} 
\arrow[bend right=20, "\mathcal R_0^{-1}" description]{d}
&
\Theta_{0,8}^2(m_0^\nu)
\arrow[bend left=15, "\mathcal R_1^0" description]{r} 
\arrow[bend right=20, "\mathcal R_0^{-1}" description]{d}
\arrow[bend left=15, "\mathcal R_{-1}^0" description]{l} 
&
\Theta_{1,8}^2\left(-\frac{t m_0^\nu}{z} \right)
\arrow[bend left=15, "\mathcal R_{-1}^0" description]{l} 
\arrow[bend right=20, "\mathcal R_0^{-1}" description]{d}
\\
\Theta_{-2,5}^1\left(-\frac{z^2 m_0^\nu}{t^2} \right)
\arrow[bend left=15, "\mathcal R_1^0" description]{r} 
\arrow[bend right=20, "\mathcal R_0^{-1}" description]{d}
\arrow[bend right=20, "\mathcal R_0^1" description]{u}
&
\Theta_{-1,5}^1\left(\frac{z m_0^\nu}{t} \right)
\arrow[bend left=15, "\mathcal R_1^0" description]{r} 
\arrow[bend left=15, "\mathcal R_{-1}^0" description]{l} 
\arrow[bend right=20, "\mathcal R_0^{-1}" description]{d}
\arrow[bend right=20, "\mathcal R_0^1" description]{u}
&
\Theta_{0,5}^1(-m_0^\nu)
\arrow[bend left=15, "\mathcal R_1^0" description]{r} 
\arrow[bend right=20, "\mathcal R_0^{-1}" description]{d} 
\arrow[bend right=20, "\mathcal R_0^1" description]{u}
\arrow[bend left=15, "\mathcal R_{-1}^0" description]{l} 
&
\Theta_{1,5}^1\left(\frac{t m_0^\nu}{z} \right)
\arrow[bend left=15, "\mathcal R_{-1}^0" description]{l} 
\arrow[bend right=20, "\mathcal R_0^{-1}" description]{d}
\arrow[bend right=20, "\mathcal R_0^1" description]{u}
\\ 
\mathbf{\Theta_{-2,2}^0\left(\frac{z^2 m_0^\nu}{t^2} \right)}
\arrow[bend left=15, "\mathcal R_1^0" description]{r} 
\arrow[bend right=20, "\mathcal R_0^{-1}" description]{d}
\arrow[bend right=20, "\mathcal R_0^1" description]{u}
&
\mathbf{\Theta_{-1,2}^0\left(-\frac{z m_0^\nu}{t} \right)}
\arrow[bend left=15, "\mathcal R_1^0" description]{r} 
\arrow[bend left=15, "\mathcal R_{-1}^0" description]{l} 
\arrow[bend right=20, "\mathcal R_0^{-1}" description]{d}
\arrow[bend right=20, "\mathcal R_0^1" description]{u}
&
\mathbf{\Theta_{0,2}^0(m_0^\nu)}
\arrow[bend left=15, "\mathcal R_1^0" description]{r} 
\arrow[bend right=20, "\mathcal R_0^1" description]{u}
\arrow[bend right=20, "\mathcal R_0^{-1}" description]{d}
\arrow[bend left=15, "\mathcal R_{-1}^0" description]{l} 
&
\mathbf{\Theta_{1,2}^0\left(-\frac{t m_0^\nu}{z} \right)}
\arrow[bend left=15, "\mathcal R_{-1}^0" description]{l} 
\arrow[bend right=20, "\mathcal R_0^1" description]{u}
\arrow[bend right=20, "\mathcal R_0^{-1}" description]{d}
\\
\Theta_{-2,0}^{-1}\left( \frac{z^2m_0^\nu}{t^3 m_{0,x}} \right)
\arrow[bend left=15, "\mathcal R_1^0" description]{r} 
\arrow[bend right=20, "\mathcal R_0^{-1}" description]{d}
\arrow[bend right=20, "\mathcal R_0^1" description]{u}
&
\Theta_{-1,0}^{-1}\left( -\frac{z m_0^\nu}{t^2 m_{0,x}} \right)
\arrow[bend left=15, "\mathcal R_1^0" description]{r} 
\arrow[bend left=15, "\mathcal R_{-1}^0" description]{l} 
\arrow[bend right=20, "\mathcal R_0^{-1}" description]{d}
\arrow[bend right=20, "\mathcal R_0^1" description]{u}
&
\Theta_{0,0}^{-1}\left( \frac{m_0^\nu}{t m_{0,x}} \right)
\arrow[bend right=20, "\mathcal R_0^1" description]{u}
\arrow[bend right=20, "\mathcal R_0^{-1}" description]{d}
\arrow[bend left=15, "\mathcal R_{-1}^0" description]{l} 
\arrow[bend left=15, "\mathcal R_1^0" description]{r} 
&
\Theta_{1,0}^{-1}\left( -\frac{m_0^\nu}{zm_{0,x}} \right)
\arrow[bend left=15, "\mathcal R_{-1}^0" description]{l} 
\arrow[bend right=20, "\mathcal R_0^{-1}" description]{d}
\arrow[bend right=20, "\mathcal R_0^1" description]{u}
\\
\Theta_{-2,0}^{-2}(C_{-2}^{-2})
\arrow[bend left=15, "\mathcal R_1^0" description]{r} 
\arrow[bend right=20, "\mathcal R_0^1" description]{u}
&
\Theta_{-1,0}^{-2}(C_{-1}^{-2})
\arrow[bend left=15, "\mathcal R_1^0" description]{r} 
\arrow[bend left=15, "\mathcal R_{-1}^0" description]{l} 
\arrow[bend right=20, "\mathcal R_0^1" description]{u}
&
\Theta_{0,0}^{-2}(C_{0}^{-2})
\arrow[bend left=15, "\mathcal R_{-1}^0" description]{l} 
\arrow[bend left=15, "\mathcal R_1^0" description]{r} 
\arrow[bend right=20, "\mathcal R_0^1" description]{u}
&
\Theta_{1,0}^{-2}(C_{1}^{-2})
\arrow[bend right=20, "\mathcal R_0^1" description]{u}
\arrow[bend left=15, "\mathcal R_{-1}^0" description]{l} 
\end{tikzcd}
\end{center}
  \caption{A part of the orbit of the seed $\tau^W_\mathcal L$-symmetry $\Theta(m_0^\nu)$}
  \label{fig:5}
\end{figure}
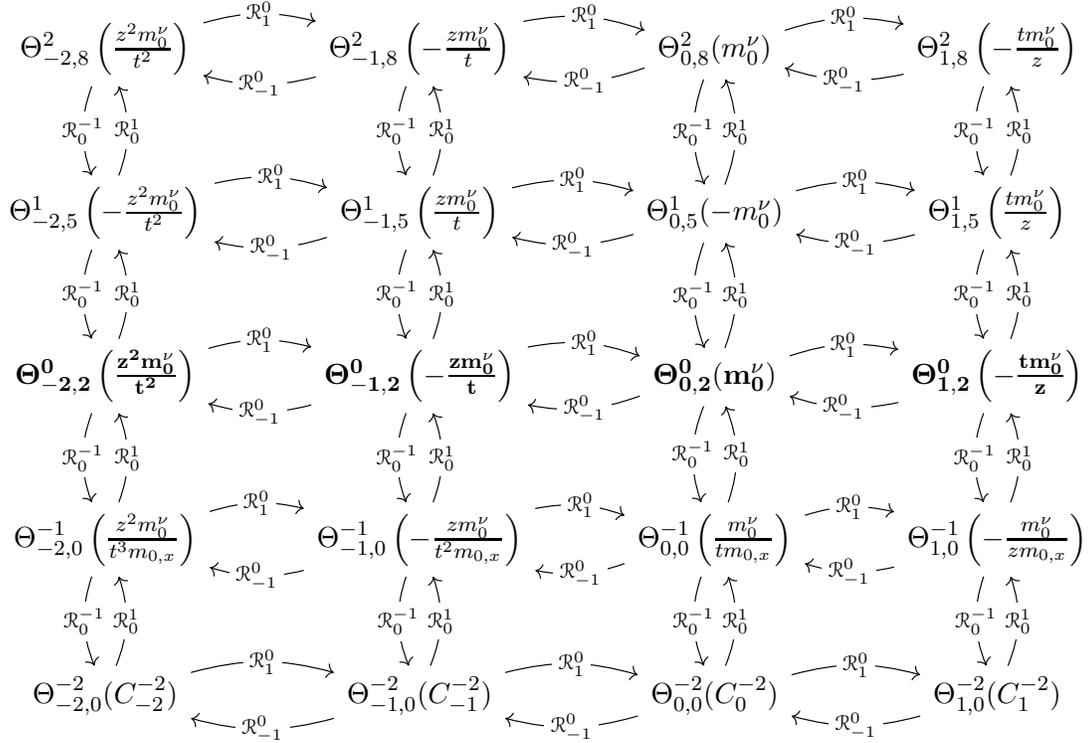

The elements of the orbit $\langle\mathcal{R}^q\rangle\cdot\Theta(m_0^{\nu})$ lie on the `horizontal line'. All these $\tau^W_\mathcal L$-symmetries are invisible ones that can be described by the following concise formula:
\begin{align*}
\Theta_{\mu}^0(&m_0^\nu)=\mathcal{R}_{\mu}^0\left(\Theta(m_0^{\nu})\right)\\
&= (-1)^{-\mu} \left[ 0,0, (z/t)^{-\mu} m_0^\nu, 0,0,\frac{1}{1 !}\mathfrak T((z/t)^{-\mu} m_0^\nu),0,0, \frac{1}{2 !}\mathfrak T^2((z/t)^{-\mu} m_0^\nu),0,0,\ldots\right].
\end{align*}
Note that all these terms belong to the vector space $\mathrm{Fam}_{\Theta}$ from Sec. \ref{fam_th}. More precisely, it can be easily seen that the equality
$\Theta_{\mu}^0(m_0^\nu)=\Theta((-t/z)^\mu m_0^{\nu})$ is true for all $\nu,\mu\in\mathbb{Z}$.
\begin{remark}
\label{orbit_theta}
In Fig. \ref{fig:5}, we present a part of the hierarchy related to the seed $\tau^W_\mathcal L$-symmetry $\Theta(m_0^\nu)$. The notation of the symmetries is again slightly modified in accordance with the rules explained in Rem. \ref{orbit_xi}. The exact forms of some of the shadows of the $\tau^W_\mathcal L$-symmetries belonging to this orbit are given as follows:

\begin{align*}
C_{-2}^{-2} &= -\frac{\nu z^2 m_1 m_0^{\nu-1} + 2zm_0^\nu}{t^3 m_{0,x}} + \frac{z^2 m_{1,x} m_0^\nu}{t^3 m_{0,x}^2},  & C_{-1}^{-2} &= \frac{\nu z m_1 m_0^{\nu-1} + m_0^\nu}{t^2 m_{0,x}} - \frac{z m_{1,x} m_0^\nu}{t^2 m_{0,x}^2},\\[2mm] 
C_0^{-2} &= -\frac{\nu m_1 m_0^{\nu-1}}{t m_{0,x}} + \frac{m_{1,x} m_0^\nu}{t m_{0,x}^2}, & C_1^{-2} &= \frac{\nu z m_1 m_0^{\nu-1} + m_0^\nu}{z^2 m_{0,x}} - \frac{m_{1,x} m_0^\nu}{z m_{0,x}^2}.
\end{align*}
\end{remark}

As already mentioned, the members $\Theta_{j,2}^0((-t/z)^jm_0^{\nu})=\Theta((-t/z)^jm_0^{\nu})$, $j\in\mathbb{Z}$, belong to the vector space $\mathrm{Fam}_{\Theta}$ from Sec. \ref{fam_th}. As usual, these symmetries are highlighted in bold and they demarcate the boundary between the area of the invisible $\tau^W_\mathcal L$-symmetries and the area of the $\tau^W_\mathcal L$-symmetries with nontrivial shadows which runs in the vertical direction this time. Thus, the crossing between these areas is realized via the basic recursion operator $\mathcal R_0^1$ and its inversion.

\begin{remark}
Since the recursion operator $\mathcal{R}_1^0$ maps the set of generators of $\mathrm{Fam}_{\Theta,k}$ to the set of generators of  $\mathrm{Fam}_{\Theta,k+1}$, we can conclude that the vector space $\mathrm{Fam}_{\Theta}$ is $\mathcal{R}^q$-invariant.
\end{remark}

\section*{Conclusions}
The main goal of our paper was to demonstrate the power of the full-fledged recursion operators in the construction of the Lie algebras of full-fledged nonlocal symmetries and subsequently in the study of particular properties of their elements. To this end, we introduced a $\mathbb{Z}$-grading on the Lie algebra of all $\tau_{\mathcal{L}}^W$-symmetries of the reduced quasi-classical self-dual Yang-Mills equation \eqref{yme}, which allowed us to study the action of the recently found \cite{jahn_voj} full-fledged recursion operators $\mathcal{R}^q$ and $\mathcal{R}^m$ in more detail. Specifically, we proved that both of these recursion operators are graded homogeneous automorphisms of the vector space $\mathrm{sym}_{\mathcal{L}}^{\tau^W}(\mathcal{E})$. Among other things, it follows from this that if we consider the action of the group $\langle\mathcal{R}^q\rangle$ on the set $\mathrm{sym}_{\mathcal{L}}^{\tau^W}(\mathcal{E})$ of all $\tau_{\mathcal{L}}^W$-symmetries of the rYME \eqref{yme}, the orbit of any non-zero homogeneous symmetry  always contains infinitely many linearly independent elements. 
Furthermore, using these results, we have described the Lie algebra $V\subseteq \mathrm{sym}_{\mathcal{L}}^{\tau^W}(\mathcal{E})$ of all known to us $\tau^W_\mathcal L$-symmetries of the rYME \eqref{yme} as the linear span of the elements of the orbits of  selected $\tau^W_\mathcal L$-symmetries (the so-called seed generators) under the action of the group $\langle\mathcal{R}^q,\mathcal{R}^m\rangle$ on the set $\mathrm{sym}_{\mathcal{L}}^{\tau^W}(\mathcal{E})$. This allows us to compute any $\tau_{\mathcal{L}}^W$-symmetry from $V$ in its full form solely by the iterative application of the formulas \eqref{ROq} - \eqref{InvROm} defining $\mathcal{R}^q$ and $\mathcal{R}^m$ and their inversions to the seed generators, followed by summation.  In this way, we can bypass the classical impractical approach, where each found shadow of a nonlocal symmetry requires additional separate calculations of the components of  its lift to the given covering.
Using the advantageous approach provided by the full-fledged recursion operators, we have also described selected members of the hierarchies related to the seed generators of the Lie algebra $V$.

Moreover, the fact that the recursion operator $\mathcal{R}^q$ is a graded automorphism of $\mathrm{sym}_{\mathcal{L}}^{\tau^W}(\mathcal{E})$  of degree -1 ensured that the union of the orbits of linearly independent elements of the zero weight is a linearly independent set.
In particular, the subset
$$\left\{\mathcal{R}_j^k(\Omega), \mathcal{R}_j^k(\Psi),\mathcal{R}_j^{k}(\Xi(q_0^\nu)),\mathcal{R}_j^k\left(\Upsilon\left(\left( \frac{y}{t^2} \right)^\nu\right)\right), \mathcal{R}_j^k(\Theta(m_0^{\nu}) \ |\ j, \nu\in\mathbb{Z}\right\}$$ 
 of the set of all generators of  the vector space $V$ is linearly independent for each fixed $k\in\mathbb{Z}$.

An open question is whether it is possible to select even larger linearly independent subsets from the set of all generators of  the vector space $V$ than those provided above.
The unresolved nature of this question is related to the fact that the recursion operator $\mathcal{R}^m$ is a graded vector space automorphism of degree zero, so, in general, we do not have guaranteed linear independence of the elements of any orbit of any homogeneous $\tau^W_\mathcal L$-symmetry under the iterative application of $\mathcal{R}^m$ and its inverse. Therefore, it would be interesting to study the orbits of the seed generators under the action of the group $\langle\mathcal{R}^m\rangle$ more closely, as well as to study the operator $\mathcal{R}^m$  itself in more detail - for example, to find its spectrum, etc. However, we leave this as an open problem for future research.

We are pretty sure that, just as with the rYME \eqref{yme}, one can construct and study Lie algebras of full-fledged nonlocal symmetries of other equations. In other words, the rYME \eqref{yme} was chosen for this study mainly because its full-fledged recursion operators are already known from our previous paper \cite{jahn_voj}. Nevertheless, the same approach can definitely be  applied to (at least some of) the four-dimensional equations from Tabs. 2-5 in \cite{DFKN}, the three-dimensional equations studied in \cite{BKMV}, or the Veronese web equation, see e.g. \cite{KMV}. Of course, as the first step, the full-fledged forms of their recursion operators have to be found, which is also an issue for forthcoming research.

\section*{Acknowledgments}
\label{sec:acknowledgments}

The symbolic computations were performed using the software \textsc{Jets} \cite{Jets}. Computational resources were provided by the e-INFRA CZ project (ID:90254), supported by the Ministry of Education, Youth and Sports of the Czech Republic. The research was supported by the Ministry of Education, Youth and Sports of the Czech Republic (MSMT CR) under RVO funding for IC47813059.


\begin{thebibliography}{99}
\bibitem{Bar} H.~Baran, \emph{Infinitely many commuting nonlocal symmetries for modified Mart\'inez Alonso-Shabat equation}, Commun. Nonlinear Sci. Numer. Simul. \textbf{96} 
(2021), Paper No. 105692, 4 pp. \href{https://doi.org/10.1016/j.cnsns.2021.105692}{DOI:10.1016/j.cnsns.2021.105692}, \href{https://arxiv.org/abs/1911.08985}{arXiv:1911.08985}
\bibitem{BKMV} H.~Baran, I.S.~Krasil{\cprime}shchik, O.I.~Morozov, and P.~Voj\v{c}\'{a}k, \emph{Nonlocal Symmetries of Integrable Linearly Degenerate Equations: A Comparative Study}, Theor.\ and Math.\ Phys., \textbf{196} (2018), Issue 2, 1089--1110. \href{https://doi.org/10.1134/S0040577918080019}{DOI:10.1134/S0040577918080019}, \href{https://arxiv.org/abs/1611.04938}{arXiv:1611.04938}.
\bibitem{Jets} H.~Baran, M.~Marvan, \emph{Jets. A software for differential calculus on jet spaces and diffieties}.  \url{http://jets.math.slu.cz}.
\bibitem{boch} A.V. Bocharov et al., \emph{Symmetries and Conservation Laws for Differential Equations of Mathematical Physics}, AMS, Providence, RI, 1999.
\bibitem{DFKN} B.~Doubrov, E.~Ferapontov, B.~Kruglikov, V.~Novikov, \emph{Integrable systems in 4D associated with sixfolds in $\mathbf{Gr}(4,6)$}, International Math. Research Notices \textbf{21} (2019), 6585--6613. \href{https://doi.org/10.1093/imrn/rnx308}{doi:10.1093/imrn/rnx308}, \href{https://arxiv.org/abs/1705.06999}{arXiv:1705.06999}
\bibitem{duar}  L.G.S. Duarte, L.A.C.P. da Mota, A.F. Rocha, \emph{Finding nonlocal Lie symmetries algorithmically}, Chaos solitons \& fractals \textbf{177} (2023). \href{https://doi.org/10.1016/j.chaos.2023.114232}{DOI:10.1016/j.chaos.2023.114232}
\bibitem{fer} E.V. Ferapontov, K.R. Khusnutdinova, \emph{Hydrodynamic reductions of multi-dimensional dispersionless PDEs: the test for integrability}. J. Math. Phys. \textbf{45} (2004), 2365--2377. \href{https://doi.org/10.1063/1.1738951}{DOI:10.1063/1.1738951},  \href{https://arxiv.org/abs/nlin/0312015}{arXiv:nlin/0312015v1}
\bibitem{jahn_voj} J. Jahnov\'{a}, P. Voj\v{c}\'{a}k, \emph{On Recursion Operators for Full-Fledged Nonlocal Symmetries of the Reduced Quasi-classical Self-dual Yang-Mills Equation}. \href{https://doi.org/10.1007/s00023-024-01425-2}{DOI:10.1007/s00023-024-01425-2}.
\bibitem{jin}  M. Jin, J.J. Yang, X.P. Xin, \emph{Nonlocal symmetries ans solutions of the multi-dimensional integrable long water wave equation}. Physice Scripta \textbf{99} (2024). \href{https://doi.org/10.1088/1402-4896/ad3382}{DOI:10.1088/1402-4896/ad3382}.
\bibitem{kramor} I.S. Krasil'shchik, O.I. Morozov, \emph{Lagrangian extensions of multi-dimensional integrable equations. I. The five-dimensional Mart\'inez Alonso-Shabat equation}. Anal. Math. Phys. \textbf{13} (2023), no. 1, 20 pp. \href{https://doi.org/10.1007/s13324-022-00763-w}{DOI:10.1007/s13324-022-00763-w}, \href{https://arxiv.org/abs/2207.07936}{ arXiv:2207.07936v1}
\bibitem{KMV} I.S. Krasil'shchik, O.I. Morozov, P. Voj\v c\'{a}k, \emph{Nonlocal symmetries, conservation laws, and recursion operators of the Veronese web equation}, J. Geom. Phys. 146 (2019),103519. \href{https://doi.org/10.1016/j.geomphys.2019.103519}{DOI:10.1016/j.geomphys.2019.103519}, \href{https://arxiv.org/abs/1902.09341}{arXiv:1902.09341v3}.
\bibitem{kra1} J. Krasil{\cprime}shchik, A.M. Verbovetsky, \emph{Geometry of jet spaces and integrable systems}, J. Geom. Phys. 61 (2011), 1633--1674, arXiv:1002.0077
\bibitem{kra2} I.S. Krasil{\cprime}shchik, A.M. Verbovetsky, R. Vitolo, \emph{The Symbolic Computation of Integrability Structures for Partial Differential Equations}, Texts \& Monographs in Symbolic Computation, Springer, Berlin (2017). \href{https://doi.org/10.1007/978-3-319-71655-8}{DOI:10.1007/978-3-319-71655-8}.
\bibitem{kra3} I.S. Krasil{\cprime}shchik, A. M. Vinogradov,  \emph{Nonlocal trends in the geometry of differential equations: symmetries, conservation laws, and B\"{a}cklund transformations. Symmetries of partial differential equations}, Part I. Acta Appl. Math.  {\bf 15}  (1989),  no. 1-2, 161--209. \href{https://doi.org/10.1007/BF00131935}{DOI:10.1007/BF00131935}
\bibitem{KrasVoj} I.S. Krasil'shchik, P. Voj\v{c}\'{a}k , \emph{On the algebra of nonlocal symmetries for the \textrm{4D} Mart\'{i}nez Alonso-Shabat equation}. J. Geom. Phys. 2011;61(9):1633-74. \href{https://doi.org/10.1016/j.geomphys.2021.104122}{DOI:10.1016/j.geomphys.2021.104122}, \href{https://arxiv.org/pdf/2008.10281.pdf}{arXiv:2008.10281v1}
\bibitem{KrugMor2016} B.S. Kruglikov, O.I. Morozov, \emph{A B\"{a}cklund transformation between the four-dimensional Mart\'{i}nez Alonso-Shabat and Ferapontov-Khusnutdinova equations}, Theor. Math. Phys. 188 (3) (2016) 1358-1360. \href{https://doi.org/10.1134/S0040577916090063}{DOI:10.1134/S0040577916090063}, \href{https://arxiv.org/abs/1502.00902}{arXiv:1502.00902v1}
\bibitem{Marvan95} M. Marvan, \emph{Another Look on Recursion Operators}, in \emph{Differential Geometry and Applications}, Proc. Conf. 393, Brno, 1995.
\bibitem{mor} O.I. Morozov, \emph{Isospectral deformation of the reduced quasi-classical self-dual Yang-Mills equation}. Differential Geom. Appl. \textbf{76} (2021), 101742, 14 pp. \href{https://doi.org/10.1016/j.difgeo.2021.101742}{DOI:10.1016/j.difgeo.2021.101742}, \href{https://arxiv.org/abs/2012.06904}{arXiv:2012.06904}
\bibitem{Mor1} O.I.~Morozov, \emph{The four-dimensional Mart\'{\i}nez Alonso-Shabat equation: differential coverings and recursion operators}. J.\ Geom.\ Phys. (2014). doi:10.1016/j.geomphys.2014.05.022. \url{arXiv:1309.4993}
\bibitem{Mor-Ser} O.I.~Morozov, A.~Sergyeyev, \emph{The four-dimensional Mart\'{\i}nez Alonso-Shabat equation: reductions and nonlocal symmetries}. J.\ of Geom.\ and Phys. \textbf{85} (2014), 40--45. \href{https://doi.org/10.1016/j.geomphys.2014.05.025}{DOI:10.1016/j.geomphys.2014.05.025},  \href{https://arxiv.org/abs/1401.7942}{arXiv:1401.7942v2}
\bibitem{Olver93} P.J. Olver. \textit{Applications of Lie Groups to Differential Equations}, 2nd edition, Springer-Verlag, 1993.
\bibitem{shag}S. Shagolshem, B. Bira, \emph{Classification of nonlocal symmetries and exact solutions for 3 x 3 Chaplygin gas equation with conservation laws}, Phys. of Fluids \textbf{35} (2023). \href{https://doi.org/10.1063/5.0151753}{DOI:10.1063/5.0151753}
\bibitem{Serg} A. Sergyeyev, \emph{A simple construction of recursion operators for multidimensional dispersionless integrable systems}. J. Math. Anal. Appl., \textbf{454} (2017), 468--80. \href{https://doi.org/10.1016/j.jmaa.2017.04.050}{DOI:10.1016/j.jmaa.2017.04.050},  \href{https://arxiv.org/abs/1501.01955}{arXiv: 1501.01955}
\bibitem{vin} Vinita, S.S. Ray, \emph{Nonlocal symmetry classifications, nonlocal symmetry reductions, exact solutions and conservation laws of one-dimensional nonlinear Vakhnenko equation}, Int. Jour. of Mod. Phys. B \textbf{36} (2022). \href{https://doi.org/10.1142/S0217979222501788}{DOI:10.1142/S0217979222501788}
\bibitem{voj} P. Voj\v c\'ak, \emph{Non-Abelian covering and new recursion operators for the 4D Mart\'inez Alonso-Shabat equation}. Commun. Nonlinear Sci. Numer. Simul. {\bf 118} (2023), Paper No. 107007, 11 pp. \href{https://doi.org/10.1016/j.cnsns.2022.107007}{DOI:10.1016/j.cnsns.2022.107007}, \href{https://arxiv.org/abs/2206.10530}{ arXiv:2206.10530v1}
\bibitem{zhang} X. Zhang, B. Wang, \emph{Lie symmetry analysis, optimal system and exact solutions for a NLPDE from the reduced quasi-classical self-dual Yang-Mills equation} Eur. Phys. J. Plus \textbf{139}, 329 (2024). \href{https://doi.org/10.1140/epjp/s13360-024-05131-0}{DOI:10.1140/epjp/s13360-024-05131-0}
\end{thebibliography}
\end{document}